\documentclass{aa}
\usepackage{graphicx}
\usepackage{txfonts}
\begin{document}
   \title {The dynamical state of A548 from XMM-Newton data: X-ray and radio connection}


   \author{L. Solovyeva \inst{1}, S. Anokhin \inst{1}, L. Feretti \inst{2},
   J.L. Sauvageot \inst{1}, R. Teyssier \inst{1},
   G. Giovannini \inst{2,4}, F. Govoni \inst{3},
   and D. Neumann \inst{1}
          }

   \offprints{L. Solovyeva,
           \\ \email{lilia.solovyeva@cea.fr}}

   \institute{CEA/DSM/DAPNIA, Service d'Astrophysique,L'Orme des Merisiers, Bat.709,
     91191 Gif-sur-Yvette, France \\
            \and INAF Istituto di Radioastronomia, via P.Gobetti 101, I-40129 Bologna, Italy \\
             \and INAF - Osservatorio Astronomico di Cagliari, Loc. Poggio dei Pini, Strada 54, 09012 Capoterra (CA), Italy\\
        \and Dipartimento di Astronomia, Universita' di Bologna, via Ranzani 1, 40127 Bologna, Italy \\
             }

   \date{Received January 15, 2008; accepted March 20, 2008 }

\abstract
    {We present a detailed study of the X-ray properties of the galaxy cluster 
Abell 548b  (z=0.04), using XMM-Newton data, and discuss
the connection between the thermal properties and the presence 
of two extended relic radio sources located at the cluster periphery. }
{We wish to analyze the dynamical state of the cluster and confirm the
presence of a major merger. We will discuss the merger effects
on the extended nonthermal emission.}
{We analyzed the X-ray brightness distribution, the surface brightness 
profiles in different directions, and the spectral properties in several
cluster regions. Moreover, to better understand the dynamical 
history of this cluster, we performed an optical analysis of the cluster
galaxies.}
{From the analysis of the temperature distribution and of the surface 
brightness profiles, we find evidence of a 
shock in the northern cluster region, 
just before the location of the two extended peripheral relics. 
From the optical analysis, we find that the cluster galaxies show
a large  $\sigma_V$, and two components are needed to fit their velocity 
distribution.
Observational results were compared with a cluster simulation.  The maps of gas temperature and density distribution from the simulation agree with 
the observational data in the case of a cluster  merger  
nearly perpendicular to the plane of the sky and in the state
after the maximum core collapse.
} 
{We conclude that we are observing a galaxy cluster in a
major merger phase,
just after the maximum core collapse. The mass ratio is about 1:2, and the merger
collision is nearly perpendicular to the plane of the sky.
A shock is present in the northern cluster region,
and it is very likely responsible for the electron reacceleration and the
magnetic field amplification that will give cause the cluster relics.
The relative position of the shock and the relics is strongly affected 
by projection effects.}

\authorrunning{L. Solovyeva, S. Anokhin, L. Feretti and J.L. Sauvageot.}
\titlerunning{ Abell 548b: study of the dynamical state}
\maketitle
%

\section{Introduction}

Cluster formation is a  still active process in the local Universe.
Galaxy clusters exhibit small-and large-scale substructures that clearly
indicate of merging activity with other clusters or groups of 
galaxies.
The X-ray observations of the thermal gas in galaxy clusters allow us to study 
the hot intra cluster medium (ICM), which is the main baryon source
in the Universe. Thanks to the XMM-Newton and Chandra observations,
we can obtain detailed temperature distributions of the ICM and understand
the influence of the cluster's formation history on the thermodynamics of the 
IGM.

The galaxy cluster Abell 548b (z =0.04) is a merging cluster with
two extended radio relics (A and B), located in a cluster peripheral region,
on the same side with respect to the cluster center, and another 
possible relic source of small size (C), near in projection to
the cluster center (\citealt{Feretti2006}).
These diffuse radio sources
have no obvious connection to cluster galaxies, but are rather associated 
with the ICM.

\begin{figure}[!ht]
\begin{center}\includegraphics[
   width=7.6cm,
   keepaspectratio]{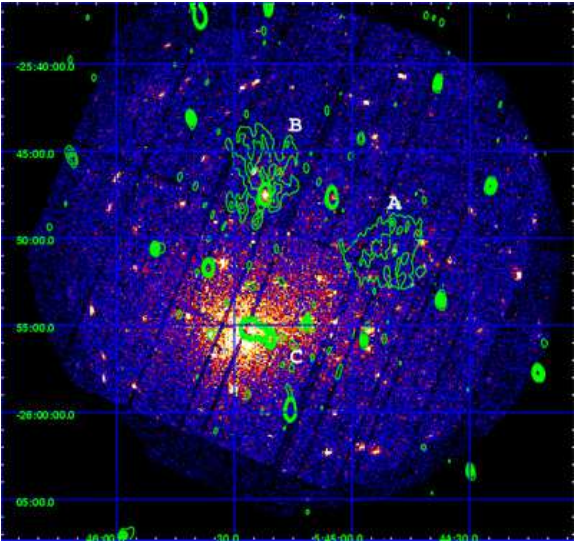}\end{center}
\caption{\label{art_uncorrected_relics} 
X-ray image from XMM data in the 0.3 - 4.5 keV energy band, 
from EPIC cameras. Green contours are the radio emission at 1.4 GHz; A, B, 
and C indicate the radio relics.}
\end{figure}

Diffuse cluster radio sources appear to fall into two 
categories: halos, which are centrally located in the cluster, relatively 
regular in shape, and unpolarized, and relics which are peripherally located, 
fairly elongated, irregular and often highly polarized.
The presence of these large regions (300-600kpc), which show a diffuse 
synchrotron emission, reveal the existence of a large-scale magnetic field
and of relativistic particles in the cluster volume.
These nonthermal
components have earned a lot of attention in the last years since their
formation and evolution is connected to the cluster's dynamical
history, and therefore they are important for a complete description
of the ICM.

Present models of the origin of relics suggest that they are tracers of merger 
shocks. Relativistic particles are accelerated by shocks either
by diffusive shock acceleration or adiabatic recompression of fossil
radio plasma (\citealt{Ensslin2002};Hoeft \& Bruggen 2007).
The production of outgoing shock waves at the cluster
periphery is indeed observed in numerical simulations of cluster
merger events (\citealt{Schindler2002}).

The cluster A548b is located in the cluster system A548, which includes 
several substructures:
A548a, A548b, and A548-2 associated with optical condensations 
(\citealt{Davis1995}). 
The combination of X-ray and optical data show that A548 is a cluster that is still in 
a formation phase and that has a complex morphology. 
The ROSAT data, presented by \citet{Davis1995} and
by \citet{Neumann1999}, indicate that this cluster is in a
merging phase, hence not yet dynamically relaxed.
A cooling core is clearly not present ( \citealt{White1997}).

In this paper we present a 
study of the dynamical state of the galaxy cluster A548b
by detailed X-ray analysis with new XMM-Newton data to understand the
connection between X-ray properties and the presence of radio relics.
In Sect. 2 we present a summary of the radio properties.
In Sect. 3 we present the new data, their reduction, and the
results of the detailed image and spectral analysis
(surface brightness profiles, 2D 
$\beta$-model analysis using XMM-Newton and ROSAT data, 
total temperature
profile, temperature profiles in different regions, 
temperature map). Section 4 gives the relationship between X-ray and 
radio data.
In Sect. 5 we compare our results with optical data and
simulations to analyze the dynamical status of the cluster.
Results are discussed in Sect. 6, and conclusions are found in Sect. 7.

\begin{figure}[!ht]
\begin{center}\includegraphics[
   width=7.6cm,
   keepaspectratio]{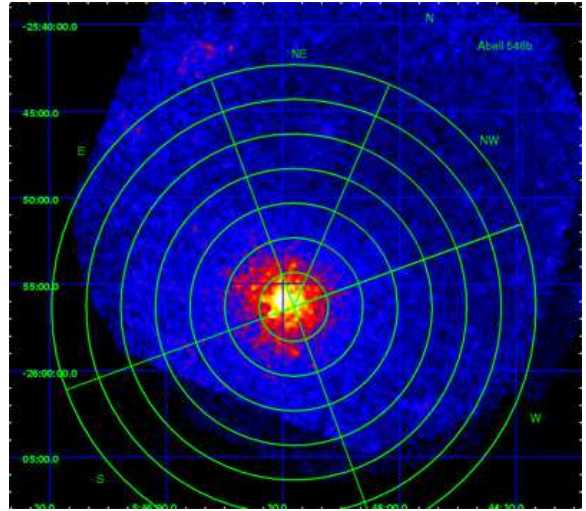}\end{center}
\caption{\label{art_corr_Panda} 
X-ray corrected image from XMM data with chosen sectors, smoothed with a Gaussian
with FWHM = 25$\arcsec$. Superimposed are sectors discussed in Sect. 3.3.3}
\end{figure}

We assume a concordance cosmology with
$H_{0}=70 $kms$^{-1}$Mpc$^{-1}$, $\Omega_{m}=0.3$,
$\Omega_{\Lambda}=0.7$. At the redshift of the Abell 548b cluster (z = 0.04), 
the angular scale of 1$\arcmin$ corresponds to a linear size of 47 kpc.

\section{Summary of radio properties}

Figure \ref{art_uncorrected_relics} shows the contours of the radio
emission at 1.4 GHz, obtained and discussed by \citet{Feretti2006}.
We concentrate here on the extended diffuse radio sources.  
A diffuse source, named A, was detected in the
western cluster region. This extended source has quite a regular
morphology and is not associated with any galaxy. Also diffuse
emission, named relic B, was detected in the northern cluster region,
around a strong point-like radio source. This extended emission is
filamentary and irregular. The embedded radio galaxy is not at the
center of the diffuse radio emission, but located at its southern
boundary. The two diffuse radio sources A and B are strikingly similar in
the total flux density at 1.4 GHz and with similar size. 
The relics A, B are
polarized at a level of 30 per cent. No
information about Faraday rotation is available. 
They have steep spectra, with a spectral index
value  of $\sim$-2$\pm$ 1.

Another apparently diffuse source (labeled C) was detected closer to
the cluster center, to the northwest, at about 1$\arcmin$ from the galaxy
triplet VV 162, classified in the literature as a dumbbell
galaxy and coincident with the X-ray centroid.
The diffuse radio source C is elongated with no obvious optical 
identification, and it is polarized at a 7\% level.
At high resolution, it shows a central diffuse emission and two brighter
and more compact spots possibly identified with cluster galaxies. 
 No sign of a radio source coincident
with the brightest cluster galaxy is detected. The total flux density of
this complex in the high-resolution image is only about 1/3 of that in
the image at lower resolution, thus indicating a diffuse
structure of lower brightness. Therefore, this source has been classified
by Feretti et  al. (2006) as a relic.

\section{X-ray analysis}

\subsection{Observations and data reduction}

Abell 548b was observed on 2006 February 17 with 70 ksec for EPIC MOS
cameras and 64 ksec for pn camera.  Figure \ref{art_uncorrected_relics}
shows the A548b observation of XMM-Newton with the radio relics
contours (A,B,C). The pointing position was chosen to be between
the cluster center and the position of radio relics to have
information on the cluster peripheral region and to study the
connection between the X-ray and radio emission.
For the data analysis we used the background of J. Nevalainen
(\citealt{Nevalainen2005}). We also checked our results using the
background of \citet{Lumb} and obtained similar
results for temperature and surface brightness profiles. We used the
method of double background substraction by \citet{Arnaud2002} (see
next section).  For the analysis, we used the XMM-Newton data
from EPIC cameras (MOS 1,2, and pn) and the XMM-Newton Science Analysis
System (SAS) for data reduction. In the MOS 1,2 data set we took into
account event patterns 0 to 12 and in pn data -- patterns 0 to 4,
flag=0. In the spectral analysis, we used the data from three EPIC cameras,
while we used only the EPIC cameras MOS2 and pn in the image analysis and 2D beta analysis, because we did not have any statistics in CCD6 for camera MOS1.
For background observation, the sky coordinates in the event files were
modified using the aspect solution of the cluster observation.  From
the data counts, we detected and excluded the periods of ``flares".
The final observing time after flare subtraction was 45 and 33 ksec
respectively.
We normalized the background with the count rate of observation in the high
energy bands  and exposure time of observation. For correction of the vignetting effect in observation
and background data, we used the weighted function
(\citealt{Majerowicz2002}).
The extended cluster emission was clearly detected, and we analyzed
the data in the 0.3-4.5 keV energy band to optimize the
signal-to-noise ratio.  Visible point-like sources were masked with
circles, and excluded from the observational data in our spatial and
spectral analysis.  We filled these gaps with the mean value of
surrounding environment by multiplying by random values from Poisson
distribution.

\begin{figure*}[!ht]
\begin{center}
\resizebox{\hsize}{!} { \hspace{0mm}
\includegraphics{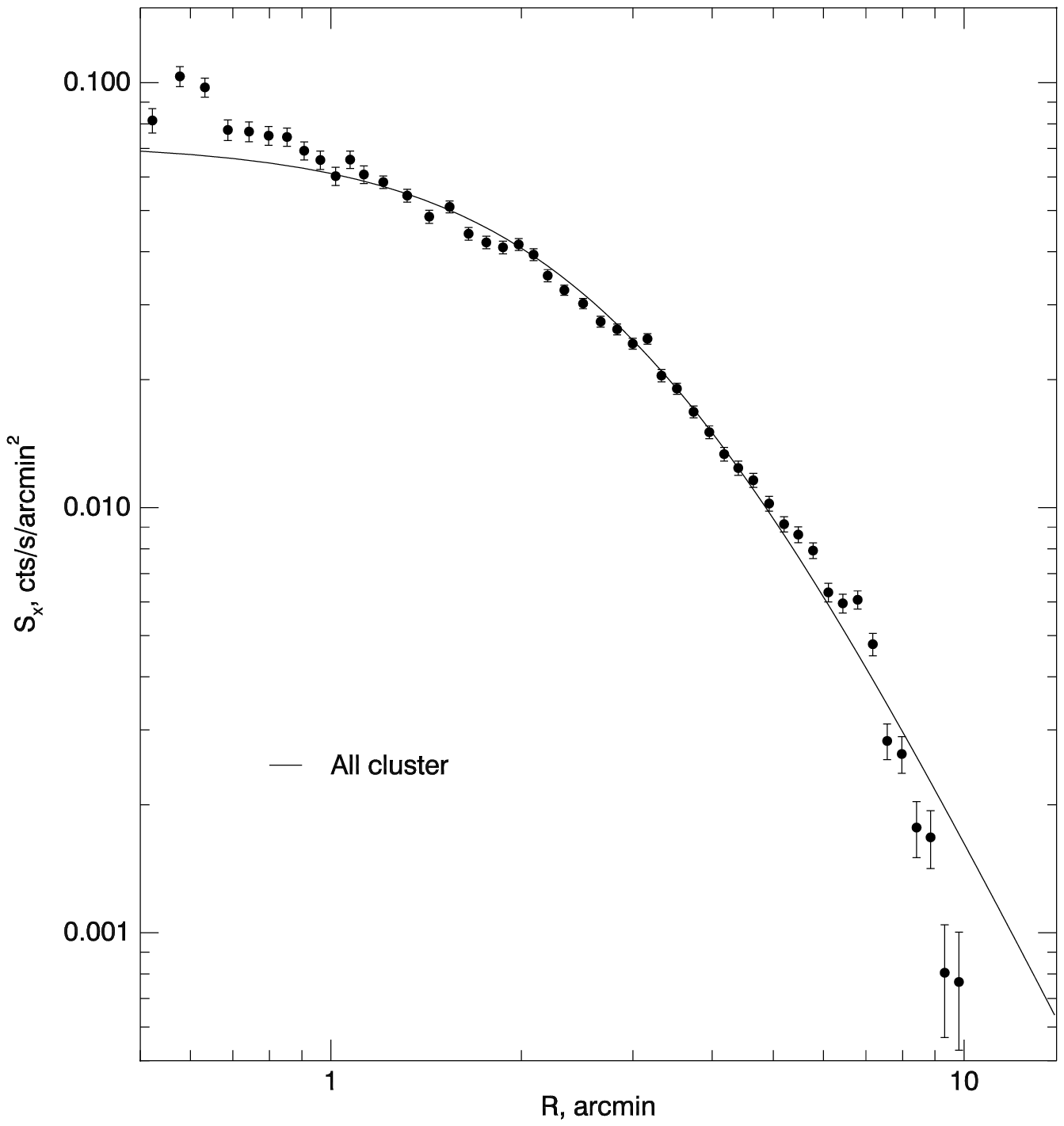}
\hspace{0mm}
\includegraphics{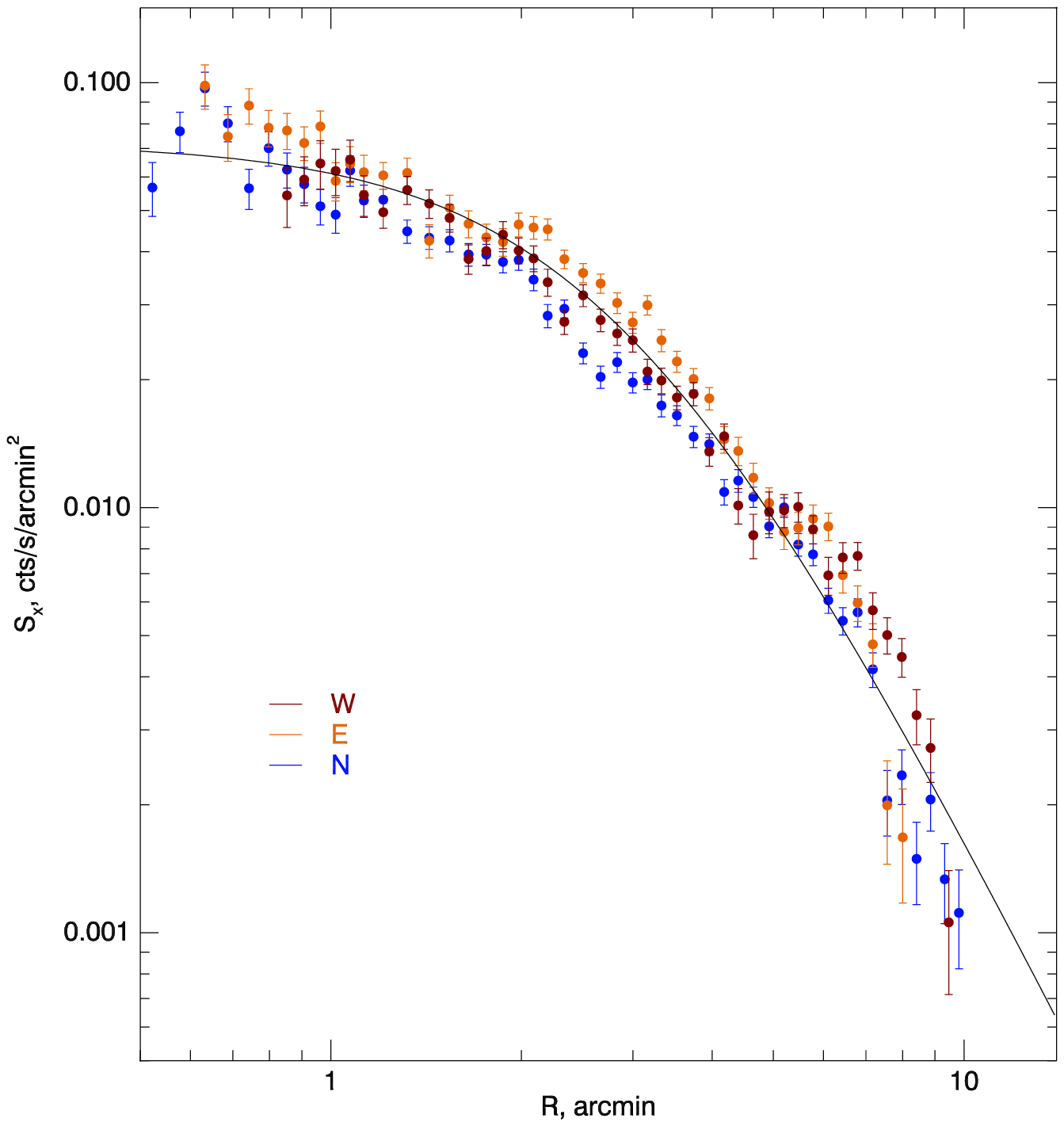}
\hspace{0mm}
\includegraphics{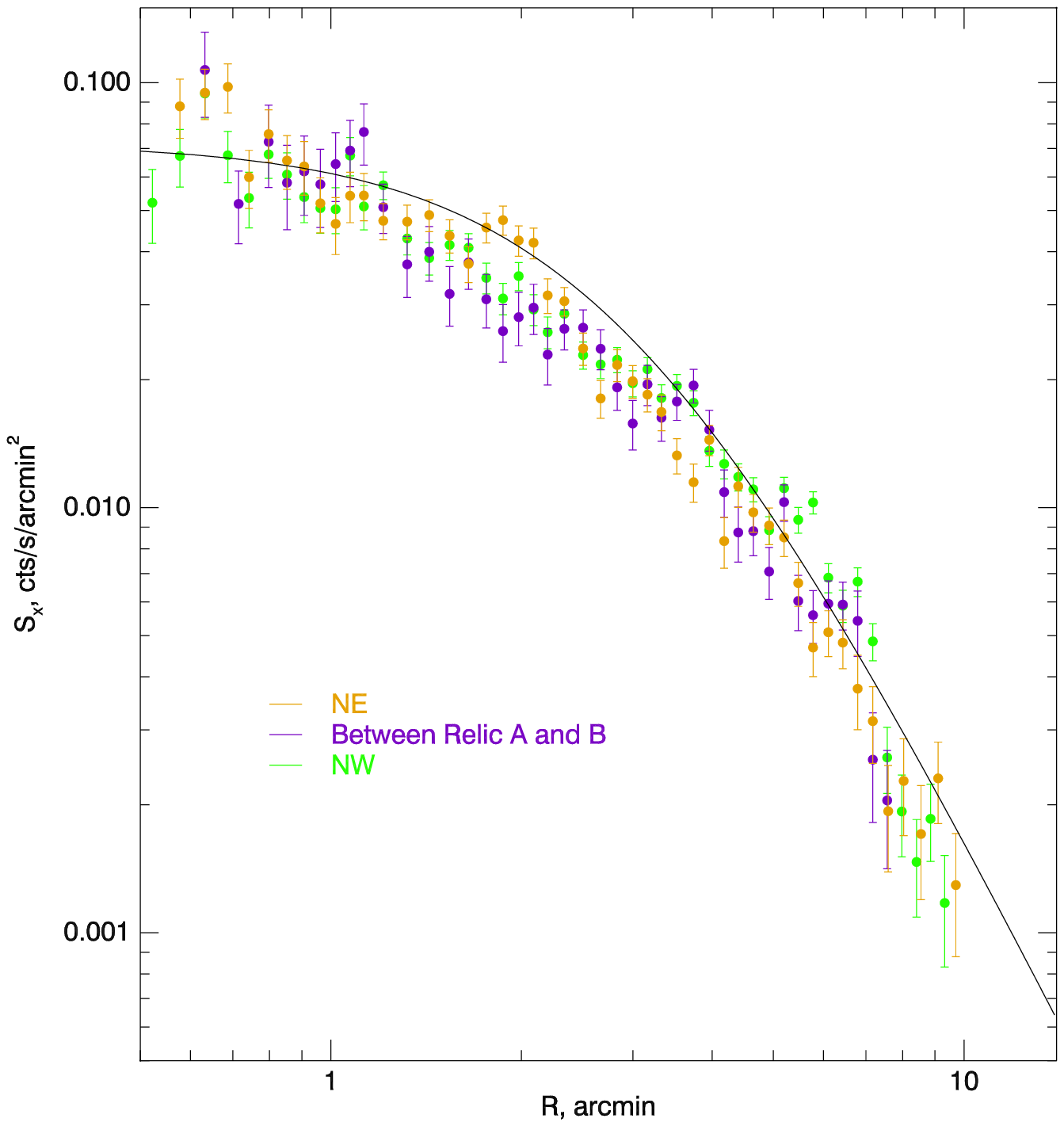}
}
 \caption{\label{art_beta} a) Total surface brightness profile
 and $\beta$-model fit. b) surface brightness profile in a different
 direction from the center cluster N, W, E and comparison with $\beta$-model fit
 obtained from radial surface brightness profile. c) surface brightness
 profile only in north direction from center cluster in particular: NE, NW and between relics A 
and B
 and comparison with $\beta$-model fit obtained from radial surface brightness 
profile.}
\end{center}
\end{figure*}

\subsection{Double background subtraction}

After cleaning  the flare events, the XMM background is dominated by
the cosmic X-ray background (CXB) and non X-ray background (NXB)
induced by high-energy particles. The background subtraction is very crucial
when reducing extended diffuse sources. 
To properly remove it, we used the
double-subtraction method by \citet{Arnaud2002}. 
It consists of two steps: (i) we subtracted the corresponding
normalized blank field obtained using the same spatial and energy
selection  (NXB component), and (ii) then
we subtracted the residual components using the data in the outer part, 
outside the cluster (CXB component).
The CXB component of the background is affected by vignetting of the
X-optics.
In the first step we used the background of \citet{Nevalainen2005}. In the second step, we minimized the
$\chi^{2}$ in the outer regions. The background counts were normalized 
using the count rate of our observation in the high-energy bands and taking the observation exposure time into account.

\subsection{X-ray brightness distribution analysis}

\subsubsection{X-ray image}

In Fig. \ref{art_uncorrected_relics} we show the A548b 
XMM-Newton image (colors) uncorrected for background or exposure. 
The superposed contour represents the radio emission. The radio relics
are named A, B, and C according to \citet{Feretti2006}.  
From this image, it is clear that the cluster 
emission it extend out to 4$\arcmin$ -- 5$\arcmin$ from the central region in
north direction (N) and up to 3$\arcmin$ -- 4$\arcmin$ in south direction (S).
The pointing position (see Sect. 3.1) is far from the X-ray peak in the
XMM image, and the cluster center is strongly affected by the vignetting effect. 
For this reason we used the position from the ROSAT image
(\citealt{Bohringer2004}) as the cluster center.

\subsubsection{Corrected X-ray image with modeling of background
 from observation data}
 
In the data analysis of extended X-ray emission from clusters of galaxies, the
background estimate and subtraction is very important. In our case we also have the problem that the vignetting effect can seriously affect our data because
of the peripheral pointing center.

For all the 2D analysis, we needed to obtain a vignetting-corrected 
and background-subtracted image. This is specially important since 
A548b was observed in the border of the field of view where vigneting effects 
is more significant. Since one component of the background, the CXB, is 
vignetted by the X-rays optics when the other, the NXB, is not, we 
need to estimate their relative contribution to be able to subtract 
them correctly. In the outer region, where the cluster contribution could 
be neglected, we can model the observation by $BKG=CXB*VF+NXB$. Since the vignetting function, $VF$, is known from calibration, one 
can disentangle the CXB from the NXB contribution by fitting this 
model to the background data in this external region. From this model 
and our outer region profile, we minimize the $\chi^{2}$ to achieve the two 
components (\citealt{Solovyeva2007}).

The obtained image of background was subtracted, and
we finally corrected our image for the gap and the presence of point sources.
We filled the regions where the mask value is zero with the mean value of the surrounding environment by multiplying by 
random values from Poisson distribution.
We then summed the images for cameras MOS2 and PN, excluding
the MOS1 camera because of the absence of CCD6.
Figure \ref{art_corr_Panda} shows the cluster image after the background
subtraction, with vignetting, gap, and point sources correction. 
This image also shows the chosen direction of sectors for study, and
the image is smoothed with a Gaussian filter 
of $\sigma$ = 7 $\arcsec$.

\subsubsection{Surface brightness profile}

Abell 548b is a non relaxed cluster (\citealt{Davis1995}), so 
we did not expect it to show a gas distribution consistent with a
single $\beta$-model.  We thus obtained fits with a $\beta$-model in
different directions. We performed a PSF convolution on the beta-model in the fit of the surface brightness profile.
In the obtained image for each XMM camera (using standard
data reduction, with the background of J. Nevalainen),  we summed and rebinned the
photons into concentric annuli with a size of $\sim$ 3.3 $\arcsec$,
centered on the center of the X-ray emission given by \citet{Bohringer2004} (see sect. 3.3.1) for each
camera. In Fig. \ref{art_beta}a we show the total radial surface brightness
profile and the obtained $\beta$-model fit.  

We also derived the surface
brightness profiles' averaging data in three sectors: N,
Western (W), and Eastern (E), defined as in Fig. \ref{art_corr_Panda}.  
The results are in Fig.  \ref{art_beta}b.  As a second
step to obtain more detailed information in the relic's direction, we
obtained surface brightness profiles in sectors NE, NW, and in between
the two relics (Fig \ref{art_beta}c).  The different sectors, except 
those between the two relics, are shown in Fig. \ref{art_corr_Panda}. 
(Note the sector between relics A and B is the angle 45-70 degree if
 the sector N is the angle 10-110 degree.)
We fitted each surface brightness profile with a $\beta$-model
(\citealt{Cavaliere1976}). We did not use the first ten points to avoid contamination from the galaxy triplet VV
162, present at the cluster cluster. 
We note that the brightest cluster galaxy is not identified with any 
discrete radio source and that no optical counterpart is associated with the 
extended emission identified as relic C (\citealt{Feretti2006} and Sect. 2).

The results of the $\beta$-model fit  are given in Table \ref{fit_beta}.
All the fits at any position angle are poor, as indicated by the high
values for the reduced $\chi^{2}$. This agrees with the fact that
A548b is a non relaxed cluster.
We also compared our results with those of \citet{Neumann1999}
obtained from ROSAT, and the comparison confirms that the region where
data show the more significant perturbation is at $\sim$
1.5$\arcmin$ and at $\sim$ 5$\arcmin$, in each of the directions for our research.

\begin{table}[!!htb]
\caption{ A548b, $\beta$-model fit in different directions }\label{fit_beta} \centering
\begin{tabular}{l c c c }
\hline
\hline
 direction & $\beta$ & $r_{c}$(arcmin) & $\chi^{2}$/d.o.f\\
\hline \hline
   total &  0.67$^{+0.08}_{-0.02}$& 3.01$^{+0.13}_{-0.20}$& 171/40\\
   West&  0.56$^{+0.03}_{-0.03}$  &2.52$^{+0.44}_{-0.45}$ & 96/33 \\
   East& 0.80$^{+0.06}_{-0.07}$ & 3.91$^{+0.40}_{-0.41}$&78/34\\
   North&0.62$^{+0.03}_{-0.03}$& 2.74$^{+0.20}_{-0.21}$& 134/40\\
\hline
North-West& 0.72$^{+0.06}_{-0.07}$& 3.79$^{+0.42}_{-0.47}$& 161/39 \\
North-East& 0.60$^{+0.03}_{-0.03}$& 2.22$^{+0.22}_{-0.25}$&65/38\\
Between  relics A B & 0.717$^{+0.10}_{-0.15}$& 3.65$^{+0.45}_{-0.47}$& 50/32\\
\hline
Neumann\&Arnaud & 0.52$\pm$0.06& 2.82$\pm$0.04&1.35 \\
from 1' to 5' & 0.54$^{+0.02}_{-0.04}$& 2.21$^{+0.14}_{-0.14}$&43/30 \\
\hline
\end{tabular}
\end{table}

\begin{figure}[!ht]
\begin{center}\includegraphics[
   width=7.6cm,
   keepaspectratio]{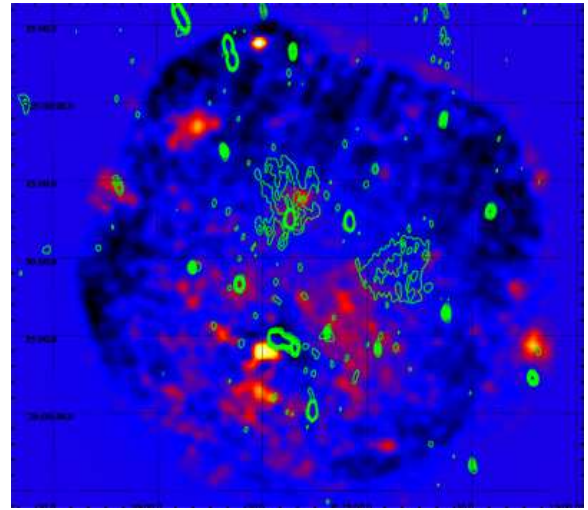}\end{center}
\caption{\label{art_Residual_Rosat_XMM.ps} 
Residual image obtained from corrected XMM-Newton image, after subtraction of the 2D $\beta$ model.}
\end{figure}

\begin{table} [!!htb]
\caption{Best-fitting results of the 2D $\beta$-model}
\label{2dbeta} \centering
\begin{tabular}{l l l}
\hline\hline
Parameter& Best-fitting \\
& XMM & Rosat\\
\hline
&\\
$R_{c1}$($\arcmin$)& 2.46 & 2.76 \\
$R_{c2}$($\arcmin$) & 2.87 &  3.34\\
$\beta$ &  0.69 & 0.65\\
Pa & 1.92 & 1.99\\
RA & 05:45:29 &  05:45:28\\
Dec &-25:55:56 & -25:55:58\\
&\\
\hline
\end{tabular}
\end{table}

\subsubsection{2D-$\beta$ analysis, X-ray residuals}

To better analyze the brightness distribution in the cluster, 
we performed a two-dimensional fit to the cluster brightness using a modified
2D- $\beta$-model that allows for two different core
radii along the two principal axes of the cluster image ellipse.
We estimated the deviations from the model, in the 
inner 6$\arcmin$.
Figure \ref{art_Residual_Rosat_XMM.ps} shows the residuals after
the model subraction from the XMM-Newton Gaussian-smoothed image.
The residual map is smoothed with a Gaussian of 25 $\arcsec$ FWHM. If one computes the significance of the structures of this size, we found roughly 10 $\sigma$ for each structure inside a 10 arcmin distance from the cluster center.

The same analysis was performed in the ROSAT data and we obtained
similar results. In particular from both data sets we obtained a very
similar residual map showing a clear decrease in the emission at
1.2$\arcmin$, and 7$\arcmin$, and an increasing emission at
4.5$\arcmin$ around the cluster center, in the relic direction (N and
W). This result agrees with the surface brightness profiles
obtained at the different directions (Sect. 3.3.2).  The best fitting
parameters are listed in Table \ref{2dbeta}. Note that XMM data give 
better results then ROSAT data since the XMM image was
obtained after a more careful subtraction of the background 
(CXB and NXB components).

The present profile analysis confirms that A548b is a 
perturbated cluster, with a complex dynamical history. 
Moreover, we observe an asymmetrical perturbation 
from the 2D- $\beta$-model displaced from the cluster center. 
To better understand the dynamical state of A548b, we need to
perform a spectral analysis now.

\subsection{Spectral analysis}

To perform the spectral analysis of the XMM-Newton data,
we used  three cameras of XMM-Newton, the background from
J. Nevalainen, and the method of double background substraction by
\citet{Arnaud2002}, and we fitted the
spectrum with XSPEC using a redshifted MEKAL plasma emission model
with absorbtion $N_{H} = 2\cdot 10^{20} cm^{-2}$ (\citealt{Davis1995}) and abundance = 0.3.
Reported errors in the text and tables are at a $68\%$ confidence level.

The best fit to the mean cluster temperature in the inner 3$\arcmin$
gives a central cluster temperature $kT=3.4\pm0.1$ keV and a reduced
$\chi^{2}$ = 1.02.  

We checked our results using models with free abundance, and obtained
similar results and, in particular, same temperature values.

\subsubsection{Temperature in different regions}

\begin{figure}[!ht]
\begin{center}\includegraphics[
   width=7.6cm,
   keepaspectratio]{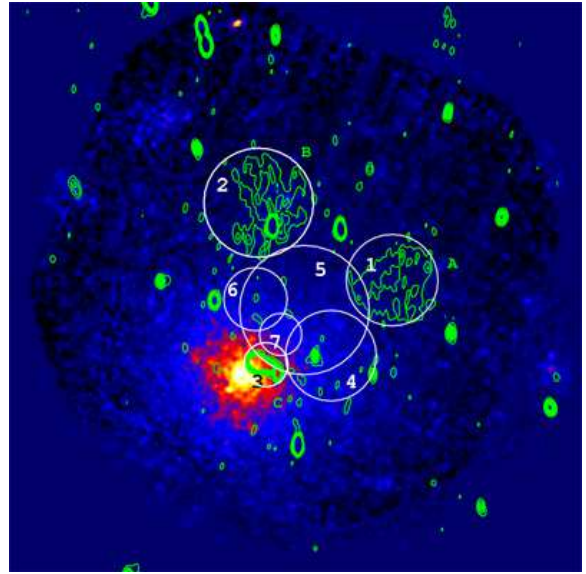}\end{center}
\caption{\label{art_RegionNum} Chosen regions of A548b for spectrum
extraction}
\end{figure}

To study the influence of the ICM gas on the radio relics, we 
extracted the X-ray spectrum in different regions near the radio relics.
Chosen regions are shown in Fig. \ref{art_RegionNum}. The results are
shown in the Table \ref{Tregions}. A temperature higher than the cluster average derived in the previous
subsection was found in regions 6 and 4, located between the
cluster center and relics A and B, with temperatures of
4.6$^{+0.7}_{-0.5}$, and 4.5$^{+0.3}_{-0.3}$ keV, respectively.  The
fits in the regions 1 and 2, i.e. in the relic regions A and B, are not
conclusive because of the poor statistics in this peripheral cluster
area.  However, in region 1 (relic A) we obtained a
temperature of 6.2$^{+3.4}_{-1.9}$ keV, suggesting a possible
temperature increase in the relic location.  To study the influence
of the ICM on relic C, we extracted the spectra in the regions 7 and
3. The obtained temperatures are 3.0$^{+0.2}_{-0.2}$keV and
3.9$^{+0.3}_{-0.3}$ keV, respectively, so very close to the cluster
average temperature in the inner 3$\arcmin$ ( 3.4$^{+0.1}_{-0.1}$
keV). 

\begin{table} [!!htb]
\caption{Spectral fits results by regions} \label{Tregions}
\centering
\begin{tabular}{c c}
\hline\hline
region & $T$(keV)$/\chi^{2}_{red}$ \\
\hline
&\\
   1 & 6.23$^{+3.4}_{-1.9}$/1.06 \\
 2 & 3.74$^{+1.4}_{-0.8}$/0.89\\
 3 & 3.95$^{+0.3}_{-0.3}$/0.80 \\
 4 & 4.46$^{+0.3}_{-0.3}$/0.88\\
 5 & 4.04$^{+0.2}_{-0.2}$/0.88\\
 6 & 4.62$^{+0.7}_{-0.5}$/0.93\\
 7 & 3.00$^{+0.2}_{-0.2}$/0.91\\
\hline
 cluster &\\
\hline
&\\
2$\arcmin$ & 3.28$^{+0.1}_{-0.1}$/0.93\\
3$\arcmin$ & 3.37$^{+0.1}_{-0.1}$/1.02\\
5$\arcmin$ & 3.59$^{+0.1}_{-0.1}$/1.09\\
&\\
\hline
\end{tabular}
\end{table}

\begin{figure}[!ht]
\begin{center}
\resizebox{\hsize}{!} {
\includegraphics[width=4.2cm, keepaspectratio]
{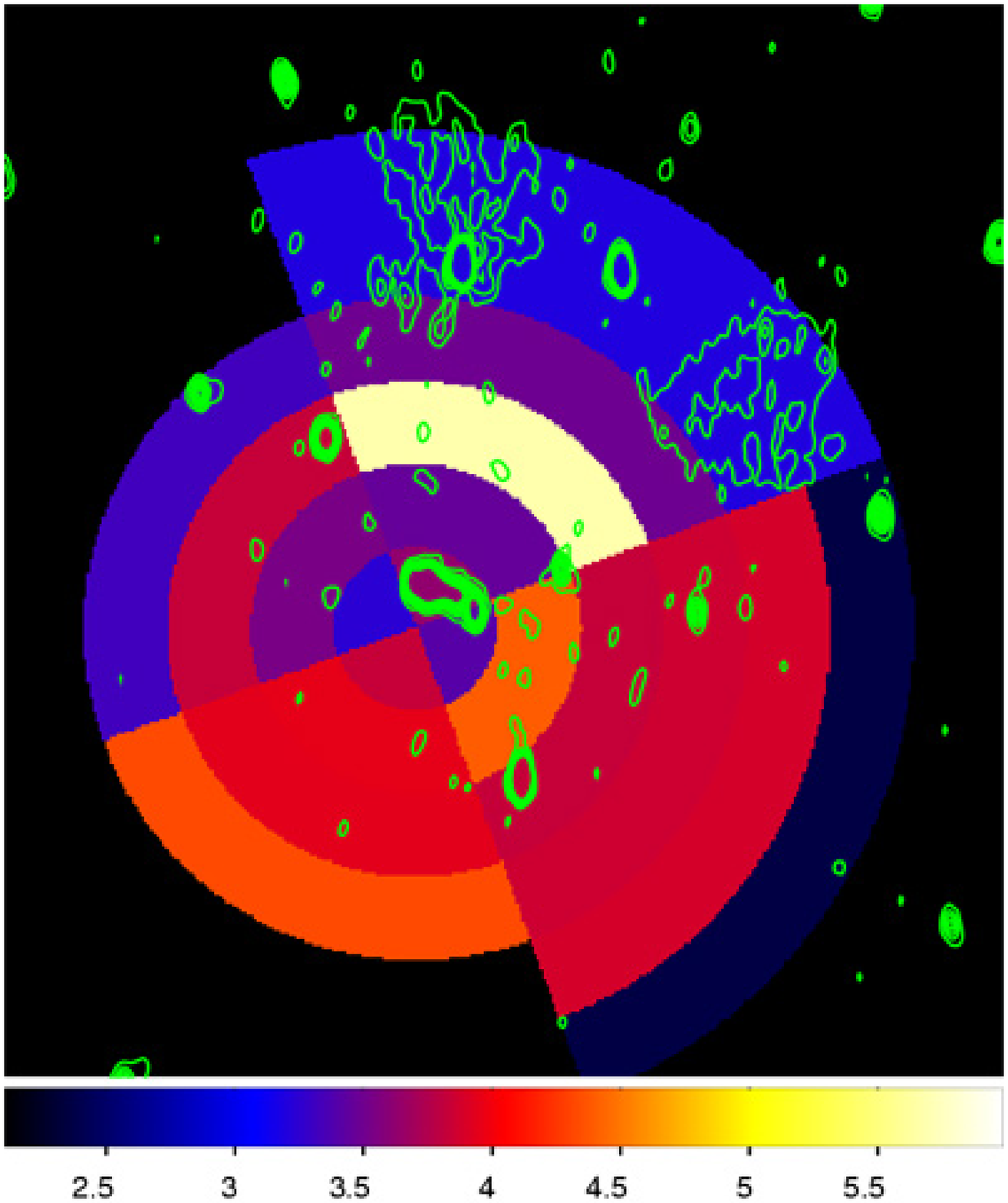}
\includegraphics[width=4.2cm, keepaspectratio]
{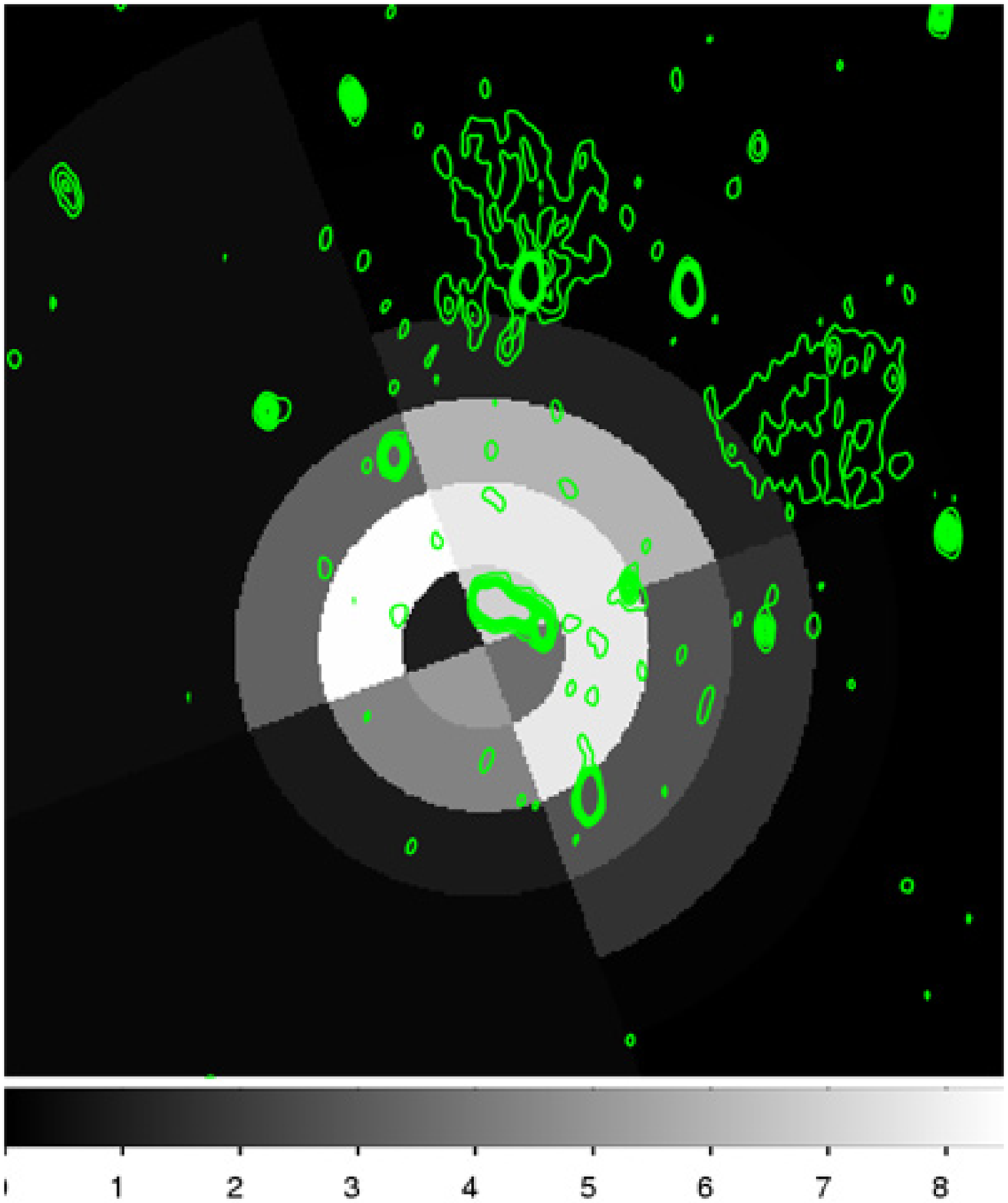}
}
 \caption{\label{art_Tmap} a)  Temperature map of A548b obtained by
sectors with contours of relics.
b) ``Significance map" = $\frac{T_{obs} - T_{average}}{T_{err}}$}
\end{center}
\end{figure}

\subsubsection{Temperature Map}

The temperature map is the best measurable indicator of equilibrium in
a system.  
We obtained temperature maps by integrating
by sectors or boxes (2$\arcmin$ in size) and by an X-ray wavelet spectral mapping 
algorithm. In all three cases we obtained similar results and find a clear 
temperature increase at $\sim$ 5$\arcmin$ from the cluster
center followed by a temperature decrease at $\sim$ 7$\arcmin$.
The most significant temperature increase was from 3.2keV to 6.5 keV,
obtained in the relic direction on the 2-3 keV.
Figure \ref{art_Tmap} we show the temperature map obtained in different sectors where
the temperature trend is more visible. In Fig.\ref{art_Tmap}b  we have the
`significance map' that shows the difference between observed and average temperature  divided on 
temperature error, where average temperature is  3.3 keV  through the whole the cluster. 

\subsubsection{Temperature profiles in sectors}

\begin{figure}[!!htb]
\begin{center}
\resizebox{\hsize}{!} { \hspace{0mm}
\includegraphics{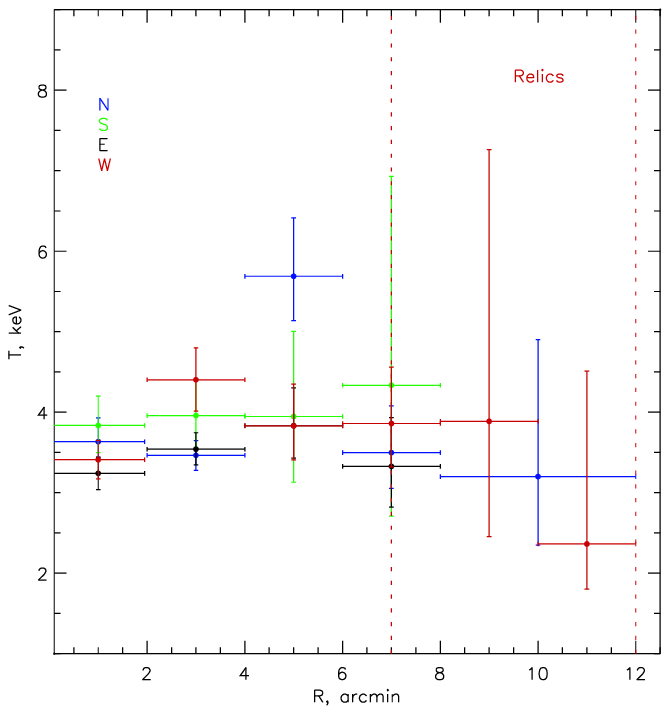}
\hspace{0mm}
\includegraphics{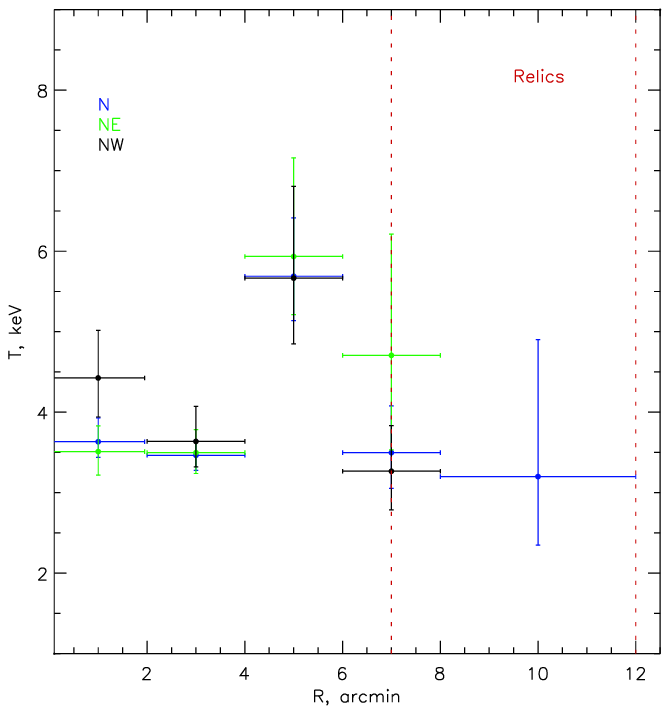}
}
 \caption{\label{art_gradT} a) temperature profiles of A548b in different directions: N, S, E, W
b) temperature profiles of A548b in relic directions: N, NW, NE.}
\end{center}
\end{figure}

For a better understanding of the previous result, we derived
temperature profiles in the different sectors for the directions: N, S,
E, W (as in Fig. \ref{art_corr_Panda}).
We extracted spectra in each sector up to 10$\arcmin$ from the
cluster center, and the size of sectors was chosen similar to the size
annuli used to derive the radial temperature profile .

Figure \ref{art_gradT}a shows obtained temperature profiles in each
direction. A significant increase in the
temperature at about 5$\arcmin$, for the N direction, (in the region just 
before relics A and B), is clear.
By splitting the N sector in the two subsectors NW and NE, 
we observed the same increase of the
temperature in the region between 4$\arcmin$ - 6$\arcmin$, before the
relic location (see Fig. \ref{art_gradT}b). 
The highest temperatures are about 5-6 keV .

\subsection{Do we detect  shocks toward the relics?}

\begin{figure}[!!htb]
\begin{center}
\includegraphics[width=\columnwidth,keepaspectratio]
{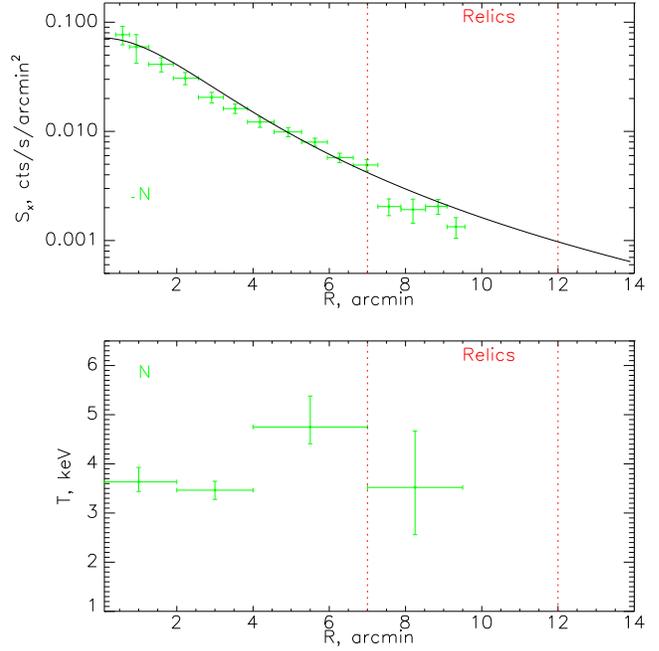}
  \caption{\label{art_gradT_beta} Obtained perturbation from X-ray data
in the surface brightness profile and in the temperature on the N
direction to relics A and B.}
\end{center}
\end{figure}

Figure \ref{art_gradT_beta}  shows the surface brightness and 
temperature profiles
in the direction of the radio relics A and B. We clearly
observe an increase at 5$\arcmin$ in the temperature profile (see Fig. \ref{art_gradT}b )
and a significative drop at 7$\arcmin$  in surface brightness, just before 
the relic location. 
On one hand, if one considers that statistics are much better in 
brightness and that surface brightness is less affected by the 
averaging along the line of sight, one should use this profile to 
estimate the shock position and its mach number.  On the other hand, 
the detection of a clear jump in the temperature could also be 
interpreted as a shock signature. That 
the projected positions of the shock are different in the two profiles 
should then be interpreted as different projection/averaging effect for surface 
brightness and temperature (also taking into account that we need a
larger bin to estimate the temperature and then suffer stronger effects).
Due to these statistical considerations, it seems better to define 
the shock position using the surface brightness profile. We, thus 
compute the temperature profile again so as to have bins just before and 
after the shock (see Fig. \ref{art_gradT_beta}).

Considering a possible shock in the N direction and taking the surface brightness profile into account(see  Fig.  \ref{art_gradT_beta}),
the density compression given by $\frac{\rho_{2}}{\rho_{1}}$ is 
1.74$^{+0.33}_{-0.27}$ at 7$\arcmin$.
We can use this value to constrain the
kinematics of the merger: the Mach number $M$ of the shock can be
derived by employing the Rankine-Hugoniot shock relations
(\citealt{Landau1959}, \citealt{Markevitch2007}) for density or 
temperature jumps of

\begin{equation}
\frac{\rho_{2}}{\rho_{1}} =\frac{(1+\gamma)M^{2}}{2+(\gamma-1)M^{2}}
\end{equation}
\begin{equation}
\frac{T_{2}}{T_{1}} =\frac{(2\gamma)M^{2}-(\gamma -1))((\gamma
-1)M^{2}+2)}{(\gamma + 1)^{2}M^{2}}
\end{equation}
where the subscript 2 denotes material behind the shock (the 
``postshock" region), and 1 denotes material ahead of the shock (the ``preshock" 
region).  For the intracluster gas, we used $\gamma = 5/3$.

Using the above compression value and Eq. 1, the data agree with
the presence of a moderately supersonic outflow with a Mach number
$M$=1.5$\pm$0.1.  This result implies a temperature jump of
$T_{2}/T_{1}$ = 1.54, in very good agreement with the measured
discontinuity, at 7$\arcmin$ : $T_{2}/T_{1}$ = 1.4$^{+0.6}_{-0.4}$.
In a first conclusion, keeping in mind both the different averaging and 
projection effect along the line of sight for temperature and density 
and the fact that both jumps lead to similar mach number, we think 
that we probably detect a relatively mild shock toward the relics.

\section{Relationship between X-ray and radio data}

\subsection{X-ray versus radio emission}

Current models propose that merger shocks are at the origin of relics.
In our case, the two radio relics A and B are well outside the cluster
core radius ($\sim$100-150 kpc), 
at projected distances of 500 and 430 kpc from
the cluster center, at the boundary of the detected X-ray 
emission.
The poor statistics do not allow us to derive conclusive information
on the temperature in the relic regions, whereas a significant
temperature increase is detected in the region before the relics.  The
presence of perturbations in the X-ray brightness distribution and of
variations in the cluster temperature 
is a clear indication of a disturbed cluster,
thus supporting the connection between the formation of relics and a
strong cluster dynamical activity. On the other hand, it seems that
there is no point-to-point coincidence between the radio regions and
the shocked region.  We show that it is possible to explain the
link between the radio emission and the X-ray properties by invoking
projection effects.

The small relic C, located at about 1$\arcmin$ from the cluster center, i.e. within 
the cluster core radius, in the NW direction, is not associated
with any obvious X-Ray substructure or temperature jump. This source 
consists of three components: a central diffuse emission and two brighter and 
more compact spots possibly associated with cluster galaxies (see Figs. 6 and 7
and the discussion in Feretti et al. 2006). The total flux density present 
in the high-resolution images is only about 1/3 of the flux density measured
in the lower resolution image (77 mJy), thus indicating the presence of a 
diffuse structure of low brightness that, according to \citet{Feretti2006} we 
associate to the relic structure. Because of the low flux density of 
discrete sources (6.9 and 4.2 mJy, respectively), the flux density and size 
of source C are only marginally affected by the presence of these possibly unrelated sources. 
This source could belong to the subclass of small-size relics
like those in A13, A85, A133, and A4038 (Giovannini and Feretti 2004,
Feretti and Giovannini 2007) located near the central brightest
cluster galaxy.  A proper study of this central region is prevented by
the presence of nearby discrete sources and the insufficent angular
resolution of the X-ray data.

\subsection{Pressure in relic regions}

The Mach number obtained in the previous subsection can be used to
estimate a theoretical pressure ratio $P_{2}/P_{1}$ $\sim$ 3.0.  This is
consistent with the observations: indeed in the direction relic A
the pressure ratio is 2.9, of relic B it is 2.3, toward the N direction it
is 2.8, and between A and B it is 2.7. In the other directions the
pressure ratio is much lower: to the W it is 1.6, and to the E it is 2.0
(see Table \ref{pressure}).  These results confirm
that the shock region is only in the relic direction. Note that to calculate 
the pressure, we used the observed electronic density in each direction
obtained from $\beta$-model fits of the surface brightness (\citealt{Neumann2005}).

We computed the minimum nonthermal energy density 
in the relic sources from the 
radio data under equipartition conditions, with standard assumptions.
We obtained about 1.6 $\times 10^{-13}$ erg cm$^{-3}$ for relics A and B and 
1.2$\times 10^{-12}$ erg cm$^{-3}$ for C. The nonthermal pressure is 9.9 
$\times 10^{-14}$ erg cm$^{-3}$ and 7.4 $\times 10^{-13}$ erg cm$^{-3}$,
respectively.
To estimate the minimum nonthermal
energy density in the source C, we used the total flux density
at low resolution, after subtracting components 7a and 7c (see Sect. 4.1). 
If we compare these results with the thermal pressure estimated in the relics
region, we note that the
minimum nonthermal pressure is about a factor 10 lower (but still in agreement
with present uncertainties) than the thermal
pressure. (From X-ray analysis we obtained 3.5$\times10^{-12}$ erg cm$^{-3}$ in the shock near relics A and B, and $\sim$ $10^{-12}$ erg cm$^{-3}$ in the relic region.) This result confirms our general analysis of the gas pressure in this region.
The higher value for relic C agrees with locat at the cluster center, and not simply projected
onto it. The equipartition magnetic field is
0.9 $\mu$G in relics A and B, and 3.6 $\mu$G in source C.

\begin{table}[!!htb]
\caption{Gas parameters in several regions, near the A, B relic location, 
inside and outside the
probable shock.}
\label{pressure}
\centering
\begin{tabular}{l l l l  }
\hline
 \hline
   & $T$& $n_{e}$& $P$\\
 region & (keV)& $(10^{-3}$cm$^{-3})$ &$(10^{-12}$dyn\ cm$^{-2})$\\
\hline
&\\
to  A, in shock&  5.7 $\pm$ 1.0& 0.40$\pm$ 0.01& 3.6 \\
to A, out shock&  3.3 $\pm$0.5  & 0.24 $\pm$ 0.01 &  1.2\\
to B, in shock&  5.9 $\pm$ 1.0& 0.33 $\pm$ 0.01&  3.2 \\
to B, out shock&  4.7 $\pm$ 1.1  & 0.19$\pm$ 0.01& 1.4 \\
N, in shock& 5.7 $\pm$ 0.6  & 0.36$\pm$ 0.01 &  3.3 \\
N, out shock& 3.5 $\pm$ 0.5 & 0.22$\pm$ 0.01 & 1.2 \\
W, in shock& 3.8 $\pm$ 0.5  & 0.39$\pm$ 0.01 &  2.4 \\
W, out shock& 3.8 $\pm$ 0.5 & 0.24$\pm$ 0.01&  1.5\\
E, in shock& 3.7 $\pm$ 0.4  & 0.45$\pm$ 0.01 & 2.6 \\
E, out shock& 3.3 $\pm$ 0.6 & 0.26$\pm$ 0.01&  1.3\\
relic A, 9' & 6.2 $\pm$ 2.2 & 0.14$\pm$ 0.01 &  1.4\\
relic B, 9' & 3.7 $\pm$ 1.2 & 0.14$\pm$ 0.01 &  0.9\\
between AB, in& 5.7 $\pm$ 0.6  & 0.37$\pm$ 0.01 & 3.4 \\
between AB, out& 3.5 $\pm$ 0.5  &0.22$\pm$ 0.01  & 1.2 \\
\hline
\end{tabular}
\end{table}

\section{Cluster dynamical state}

A548b is a cluster with complex dynamical state.  Observational
evidence suggests the presence of a shock structure before the radio
relics A and B. To explain this result, we need to understand
the cluster's dynamical history better since radio relics would generally be
expected to coincide with the shock region.

We therefore performed a study of the geometry of the merger from  
optical analysis and from simulations.
In this study, we need to take into account the projection effects.

\begin{figure}
\vspace{-0.25cm}
\includegraphics[angle=90,keepaspectratio,width=\columnwidth] 
{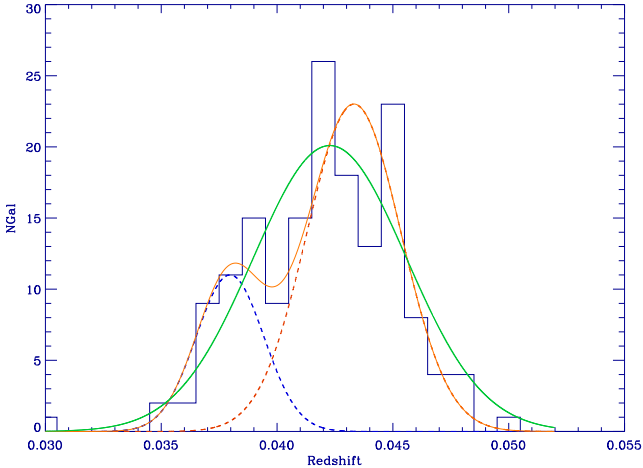}
\caption{\label{fig:NED_Histo} Histogram of the NED galaxies with known velocity at a projected distance less than $R_{200}$ from the X-ray
center.The light green line is the Gaussian fit.  A large
$\sigma_V$ = 1300 kms$^{-1}$ is necessary to fit a single Gaussian to
the data. If one tries with two Gaussians (red continuum and dashed  
lines), the $\sigma_V$ of the major component drops to $\approx$ 925
kms$^{-1}$ and
the two velocity centers found are very close to the velocity of A548-2
and A548a} 
\end{figure}

\subsection{A548b galaxies velocities and environment}

To estimate possible projection effects, we
retrieved optical information on the galaxies of A548
from the NED database. Among
606 galaxies found in NED near the A548b cluster, 371 have a velocity
measurement. Among these galaxies, 193 lie at a projected distance less than $R_{200}\approx$1.5Mpc (i.e. r$<$30.5$\arcmin$) from the A548b
cluster X-ray center. In Fig. \ref{fig:NED_Histo}, we present the
histogram of the velocities of
these galaxies. First, we obtained a fit to this
histogram using a single Gaussian. The fit result implies a large
$\sigma _V$ value of nearly 1300kms$^{-1}$.
Using this value, the expected cluster average
temperature should be $kT \sim$ 8 keV more than twice
the observed value in the relation
$kT={(\frac{\sigma_V}{323.6})}^{1.49}$ (\citealt{Xue2000}). Therefore, even if statistics are relatively
poor, we have tried to fit the galaxy velocity histogram with two  
Gaussians.
In this case, we found the major component with $\sigma_V \simeq~$ 925 
$\pm$135kms$^{-1}$ and $z$ = 0.0432,
whereas the second component has $\sigma_V\simeq~$664$\pm200$kms$^{-1}$  
and $z$ = 0.038.

Remembering that A548b (05h45m27.9s -25d56m19s V = 12708 kms$^{-1}$
z = 0.042389) belong to a structure containing A548-2 (05h42m18.0s
-26d05m00s V = 12861 kms$^{-1}$ z = 0.0429), probably in the same plane  
of the sky and A548a (05h48m38.9s -25d28m12s V = 11834 kms$^{-1}$ z = 0.039474),  
some 1000 km s$^{-1}$ less than the two others, the velocities found in the fit  
with two Gaussians, compare quite well with these redshifts
(see Fig. \ref{fig:NED_Histo}).

\begin{figure*}[!!htb]
\begin{center}
\includegraphics[width=5.5cm, keepaspectratio]
{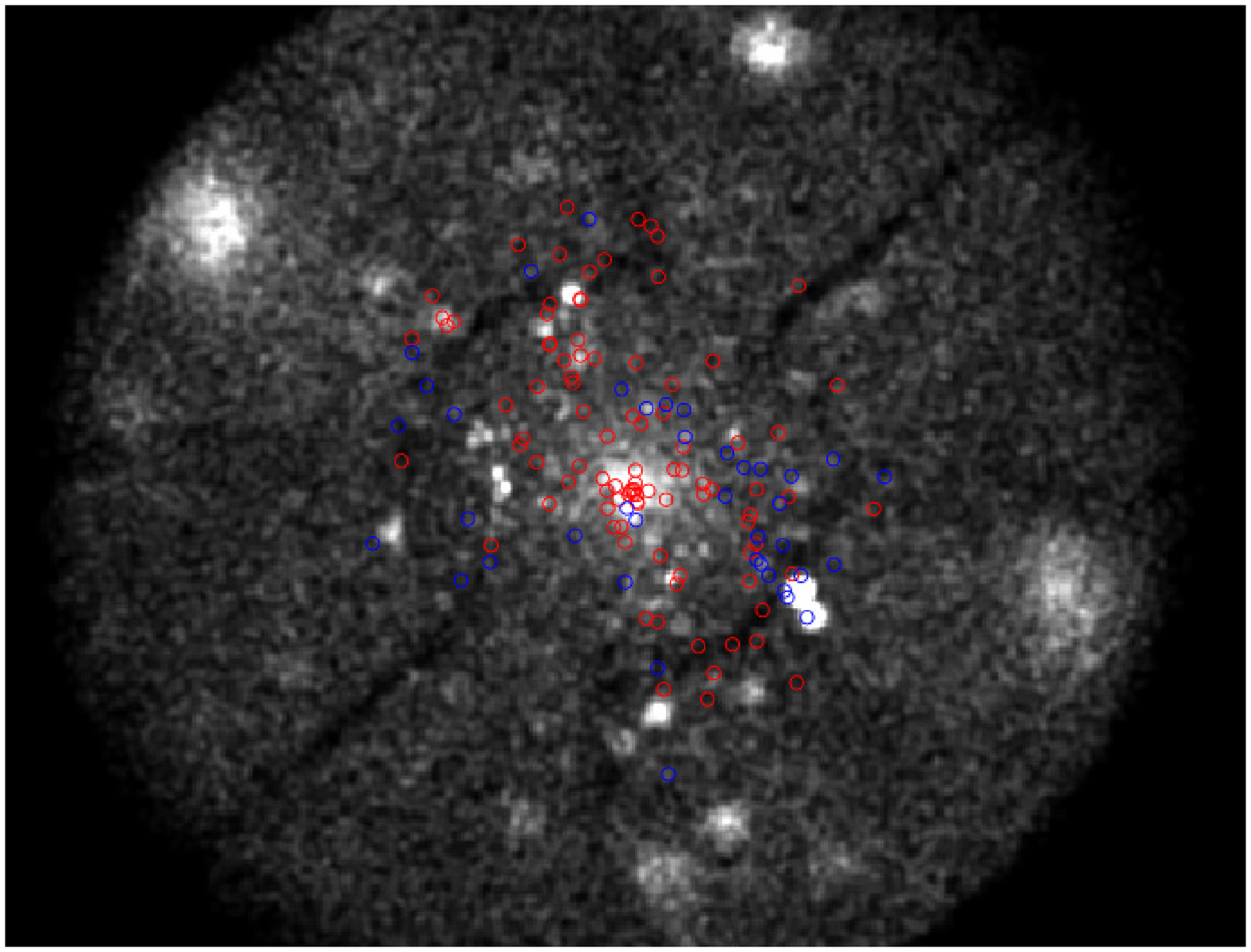}
\includegraphics[width=5.5cm, keepaspectratio]
{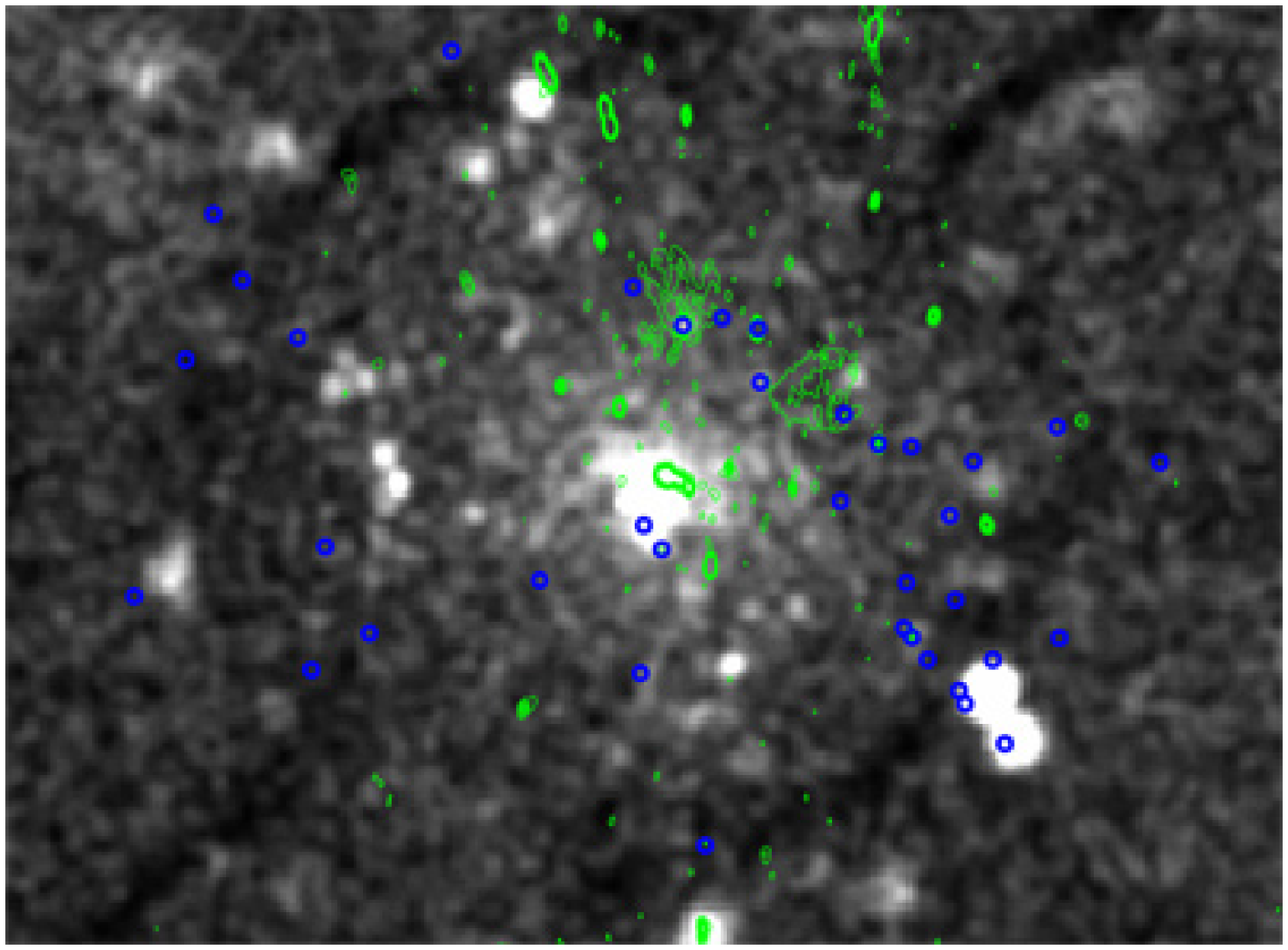}
\includegraphics[width=5.5cm, keepaspectratio]
{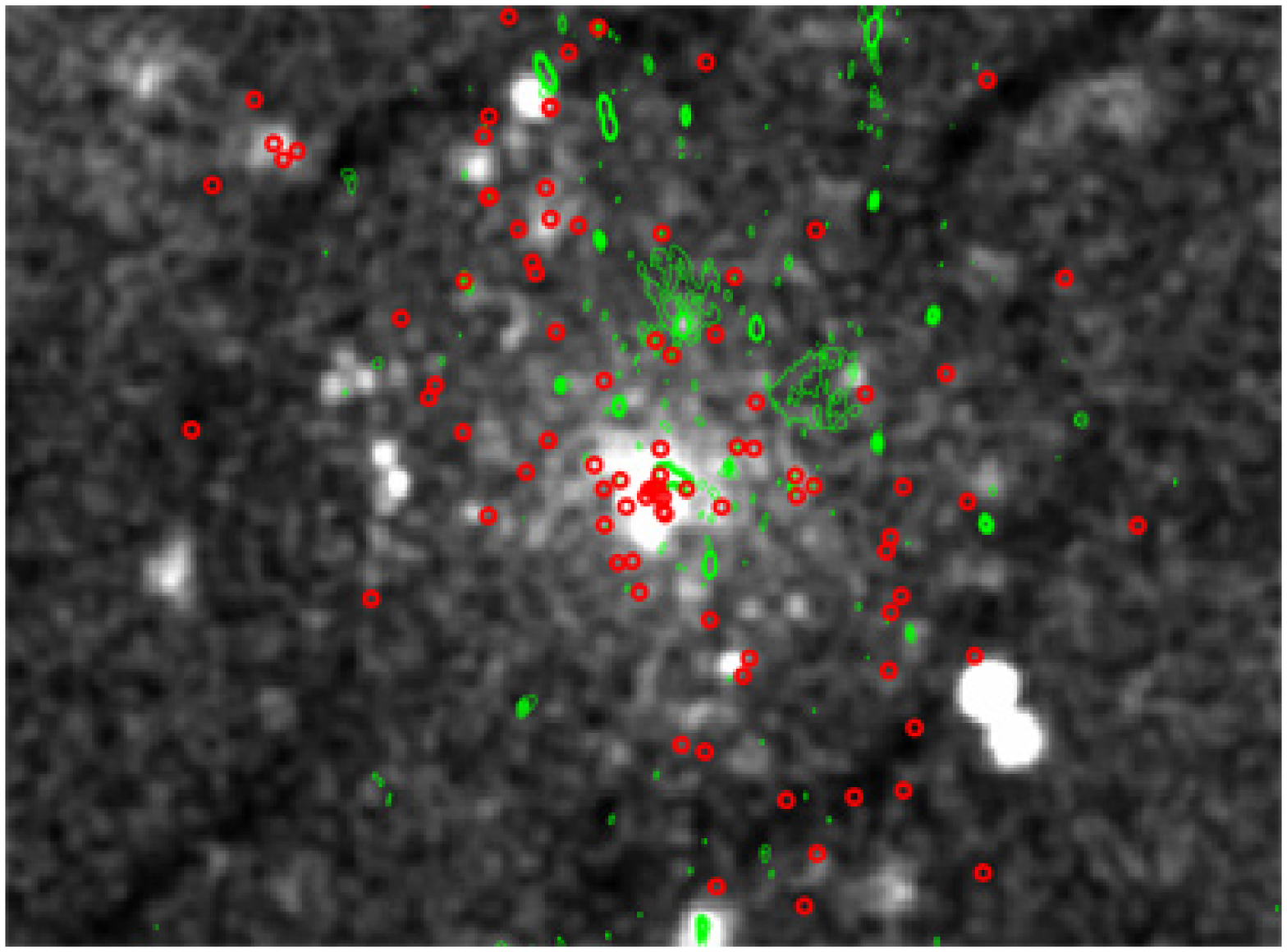}
\end{center} %
\caption{\label{fig:NED_gBR} a) X-ray from ROSAT data and
distribution of galaxies, the positions of the galaxies around the
X-ray center. b) X-ray, radio data and ``blue'' circle correspond to
galaxies having $z<0.04$. c)X-ray, radio data, and ``red'' circles
correspond to galaxies with $z>0.042$.}
\end{figure*}

We then divided the galaxy data set in two groups at a redshift  
limit of
0.04. We have 39 galaxies with $z<0.04$ and 96 galaxies with
$z>0.042$. In an Ra-Dec plane centered on the X-ray center, we plotted (Fig. \ref{fig:NED_gBR}a) these two velocities
groups as blue and red circles. One can see that the ``red''
group gently fills the whole space, while the ``blue'' one seems to be
more distributed along the NE -- SW direction.

We now try to estimate the mass ratio of the merging units. On one  
hand, if we split the histogram in two parts at $z\simeq$0.04-0.042 and consider identical galaxy mass distributions for the``blue" and ``red" samples, then the number ratio is a first approximation of the mass ratio. In this way, we find $\simeq$2.46.
On the other hand, if we use the fit results, the ratio between the areas  
of each component gives an estimate of $\simeq$ 2.70$\pm$0.94 or, if  
we try with $\sigma_V$, and $M_1/M_2\simeq(\frac{\sigma_{V1}}{\sigma_{V2}})^2$, 
we find a mass ratio $\simeq$ 1.9$\pm$1.3. If we  
remember the poorf statistics, these three estimates are  
fairly coherent even if not strongly constraining.

We can thus explain the apparently large $\sigma_V$ and the
spatial distribution of the ``red'' galaxies if we consider the
following merging scheme: A548b would be the merger of two units, the
major component being A548b, and a smaller one from the direction
of A548a.  Notice that, in this scheme, the collision axis forms a
relatively small angle with respect to the line of sight.

Different situations are possible as a merger between A548b and A548a, 
from the NE or  clump arriving from A548-2 to A548b but it is very probably so that
the merger collision is nearly perpendicular to the plane of the sky.

In conclusion, from this optical analysis, we propose that we are  
observing a merger between two clusters with $\simeq~$1:2-3 mass
ratio (estimated from the velocity histogram of galaxies as shown in
Fig. \ref{fig:NED_Histo}).
The merger's collision axis is nearly on the line of sight.
The smaller cluster probably is located in the SW from the cluster
emission center and we are observing the merger just after the core collapse.
It should be noted that we detected no galaxy associated with
the second component (``blue'' circles), behind the radio relics A  
and B, and in the cluster
center (see Fig. \ref{fig:NED_gBR}).

\subsection{Comparison with simulations}

To  understand the  dynamical  history  of A548b  and  to explain  the
connection between X-ray, radio, and optical data, we found it useful to
compare  observational  data  to  the  theoretical  predictions  of  a
high--resolution  cosmological simulation. For  that purpose,  we 
considered the 6th most massive  halo (hence the name Cluster6) in the
$z=0$  snapshot of  a dark matter only  simulation of  a $\Lambda$CDM
universe  with a periodic  box size  of 80  $h^{-1}$Mpc. We then
resimulated the same box this time using the N body and hydrodynamics
AMR code RAMSES (\citealt{Teyssier}), with a much higher resolution in
a region  of 20 h$^{-1}$Mpc around  the cluster, and  with a  much lower
resolution outside this  radius, just enough to get  the correct large-scale tidal  field. This now  rather standard simulation  technique is
usually called ``zoom simulation'' or ``re-simulation'', and it is used
in cosmology to focus computational resources on one specific object.
Cluster6 is very interesting for this paper,  since its final Virial
temperature  is  close  to  3.3 keV  and  $R_{200}\approx$1.42Mpc,  in
striking  agreement   with  A528b.   Last  but   not  least,  Cluster6
experienced a major merger around $z \simeq 0.7$, and at this redshift
Cluster6 should be close to the assumed dynamical state of A548b.  The
simulation was initialized using a $512^3$ Cartesian grid to cover the
whole box, and  then derefined outside the central  20 h$^{-1}$ sphere.
We then progressively and  dynamically increased  the  level of
refinement  {\it  only  in   the  high-resolution  region},  using  a
quasi-Lagrangian  refinement  strategy, until  we  reached 7  additional
levels of  refinement.  Our finest cell  size is therefore  close to 1
$h^{-1}$kpc,    comfortabily    smaller    than   the    observational
resolution. The  number of  AMR cells in  the Virial radius  is almost
$1\times 10^6$, while the number  of dark matter particles is slightly
above $7\times 10^5$.

\begin{figure}[!!htb]
\begin{center}
\includegraphics[width=4.2cm, keepaspectratio]
{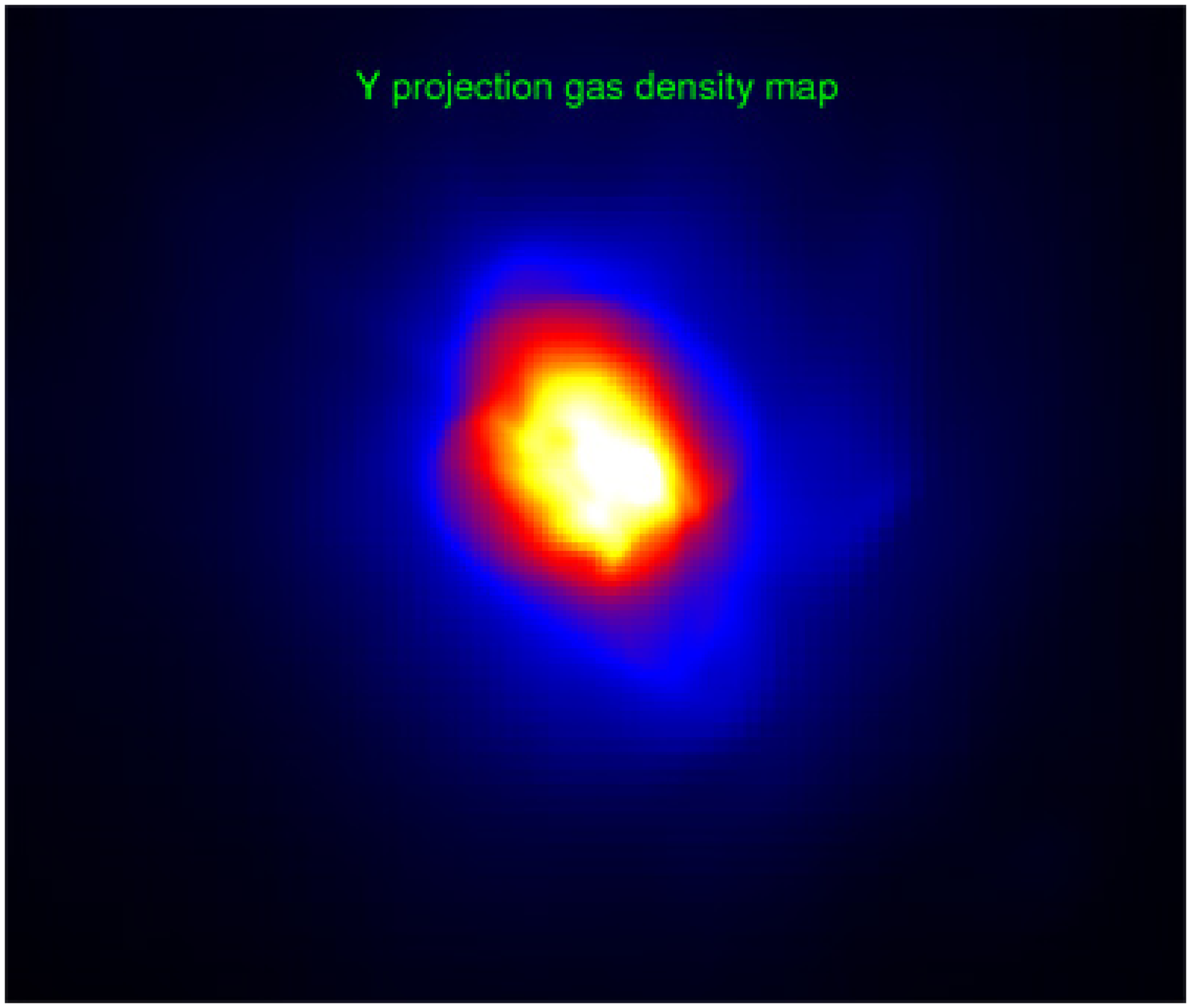}
\includegraphics[width=4.2cm, keepaspectratio]
{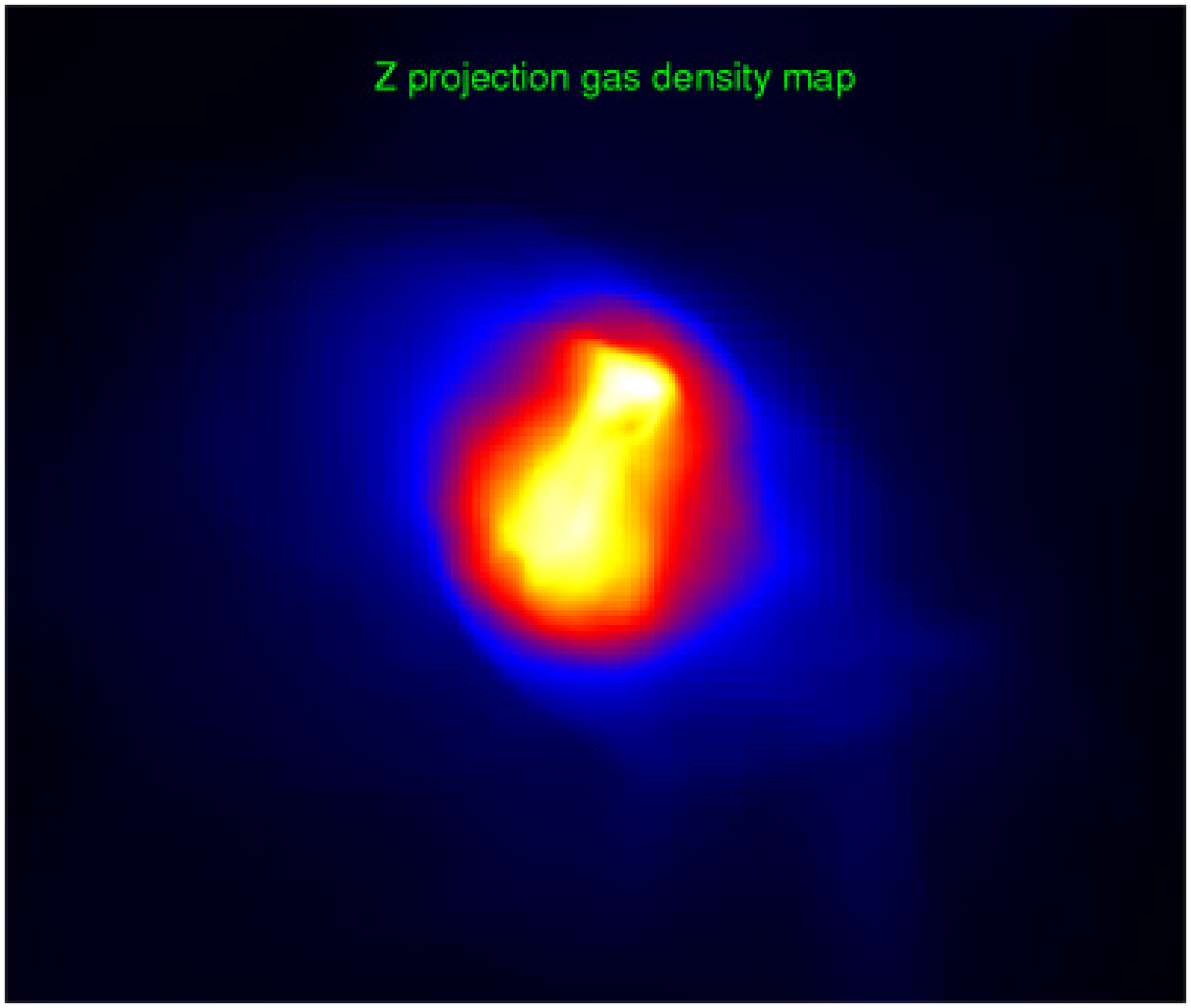}\\
\includegraphics[width=4.2cm, keepaspectratio]
{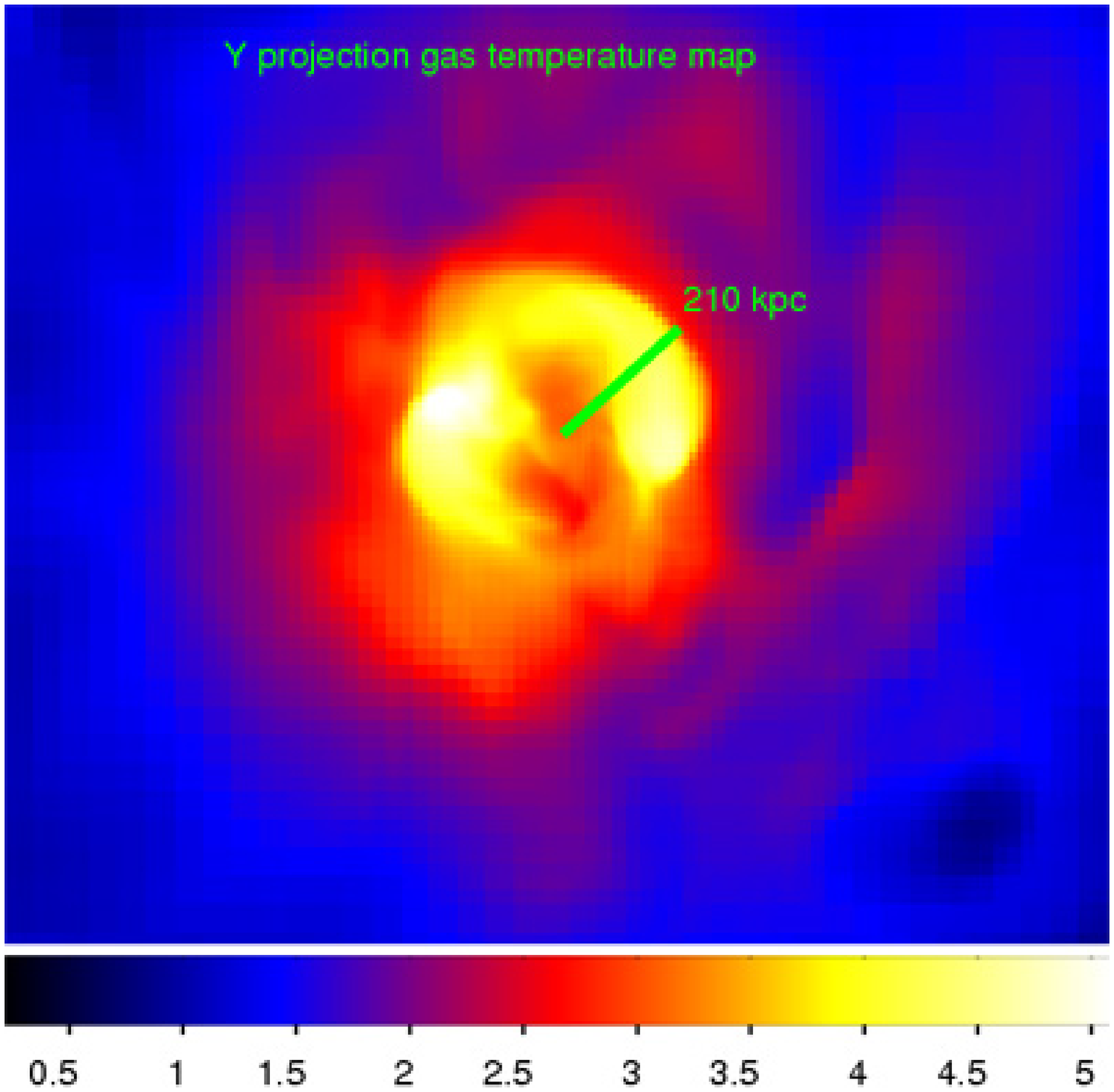} 
\includegraphics[width=4.2cm, keepaspectratio]
{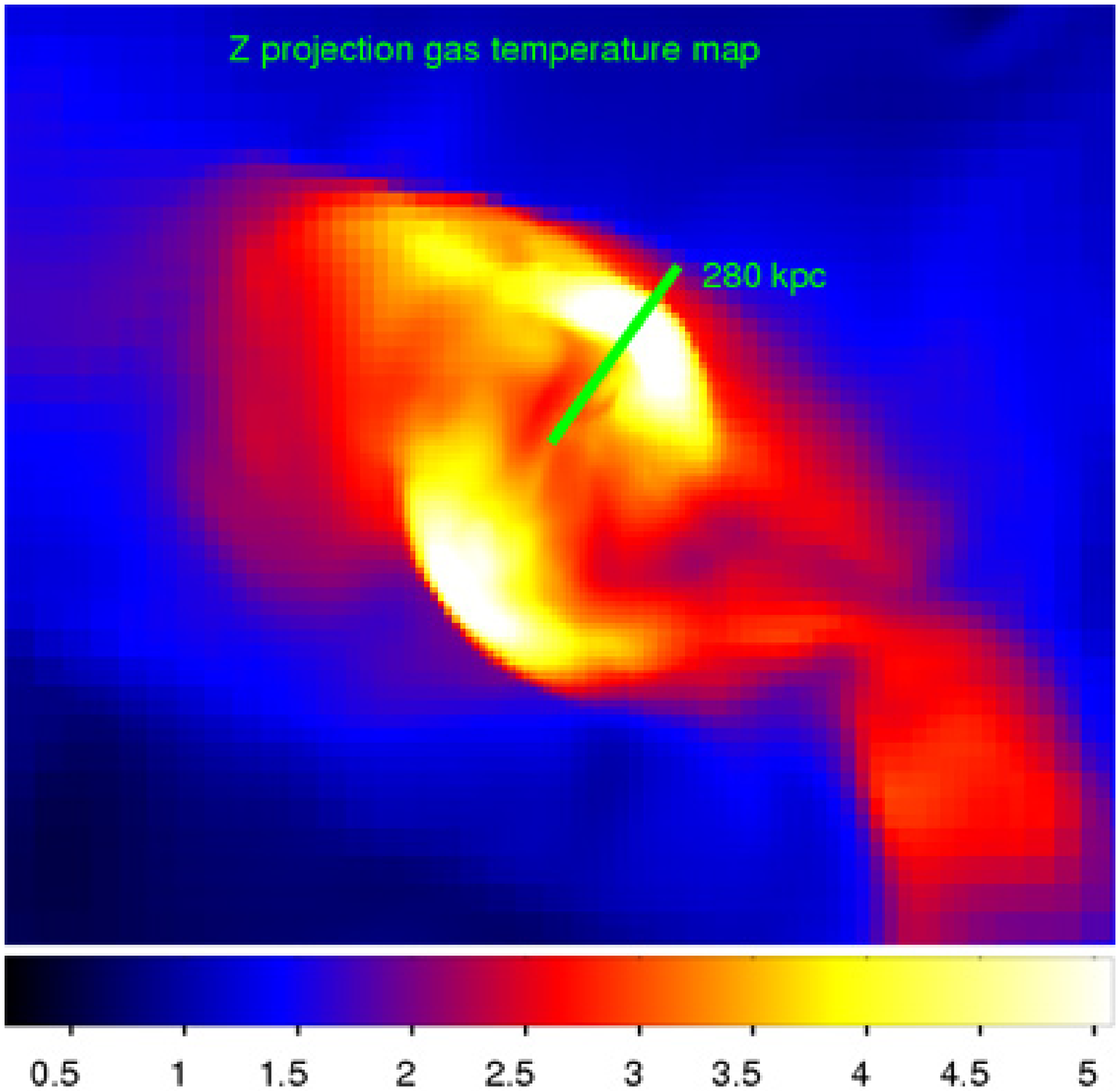}
\end{center} %
\caption{\label{simul_gas_temp} Simulation data. We
treated outputs just after maximum core collapse, in different
projections. The figure shows the same output for Y projection and for Z
projection, in particular Y gas density map, Y gas temperature map with
color scale, Z gas density map, Z gas temperature map with color scale. We observed the shock in
210 kpc with ratio T2/T1 = 2 in Y projection.}
\end{figure}

We performed the comparison
between  simulation  results  and  X-ray  data for  gas  density,  gas
temperature and, Mach number. These parameters are  important for tracing
the connection with radio data and particle acceleration.

We chose the different outputs
obtained from simulation around the maximum core collapse. We focused
on the epoch when two clumps merge along a large-scale filament and
consider different projection axes. At the same time, we studied the
effect of projection on the gas distribution. We examined the different
projection  X, Y, Z of the merger to have a complete scenario of all the
merger situations. We obtained 
different results for the gas density, gas temperature map with a the same
outputs, from the same cluster at the same time. After our analysis of different
outputs from simulation and comparison with X-ray data of A548b, we
selected the output results just after the maximum core collapse. In this
case simulated results for the temperature ratio  and distance to the shock 
agree with observational results of A548b.

Figure \ref{simul_gas_temp} shows the temperature and density maps obtained
from the Y and Z projections. The results are very different, and we select
the Y projection, because the density map has the same
distribution of the A548b emissivity map. Moreover, the temperature map 
obtained from the simulation
has a temperature increase around the cluster center, similar to the observed one.
The maximum of the temperature is in the annulus at 210 kpc, with values 
between 4.5 and 6 keV.
The same temperature of A548b (3.3 keV) was obtained at the cluster center.
Two peaks in temperature (6 keV) are present in the two regions around the 
cluster center. Note that the position of the peaks depends on the line-of-sight direction.
After the optical analysis, we derived that
A548b is in the state of merging, with 1:2-3 mass ratio of the units, 
and the merger collision 
direction nearly perpendicular to the plane of the sky.

\begin{figure}[!!htb]
\begin{center}\includegraphics[width=\columnwidth, keepaspectratio]
{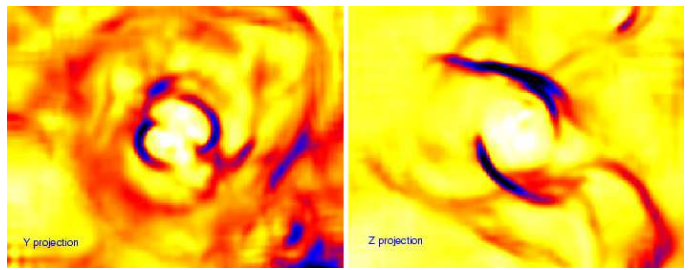}\end{center} 
\caption{\label{simul_tmap_shock} Obtained Mach number from
simulation at the same output, just after maximum core collapse for the Y
and Z projection. }
\end{figure}

To compare our results with radio data, we calculated  the Mach number map 
for different projections of chosen outputs, using the projected velocity distribution 
and the projected speed of sound, in particular (\citealt{Vazza2008}):

\begin{equation}
\Delta v = \frac{3}{4}v_{s}\frac{1-M^{2}}{M^{2}}.
\end{equation}
Figure \ref{simul_tmap_shock} shows the obtained Mach number maps for Y, Z projection, just
after maximum core collapse. In Y projection we observed a circular
shock after temperature shock in the radius 250kpc, where the
relativistic particles can be accelerated.

\section{Discussion}

\begin{figure}[!!htb]
\begin{center}
\includegraphics[width=7.6cm,keepaspectratio]{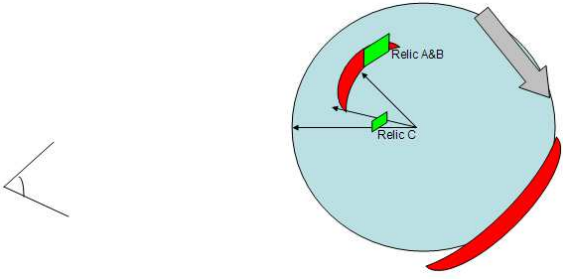}\end{center}
\caption{\label{projection}Probable geometry and merger collision
scenario. Red regions indicate the shocks, green regions indicate 
the relics. The arrow indicates the merger direction.}
\end{figure}

To explain the observational properties of A548b we need to understand the
cluster dynamical history, in particular the merger collision
scenario. After our analysis we found the following main facts:
\begin{itemize}
\item From the X-ray analysis, we observed the shock in the ICM
temperature and emission just before the relic's location, and the increase in the temperature and
highly perturbed emissivity profiles in all cluster sectors.
\item From the optical analysis, we found that we very likely observed a
merger of two units with 1:2-3 mass ratio, with the smaller unit located
in the SW.
\item From our simulations and from A548b environments, we found 
that the merger collision should be nearly perpendicular to
the plane of the sky.
\end{itemize}

To reconstruct the geometry of the merger, we compared the
observations with simulations.  
We propose the following merging scheme: A548b would be the merger of
two units, with a 1:2-3 mass ratio, a major component coming from A548b,
and a smaller one coming from the NE direction ( i.e. from A548a).
The merger collision is nearly perpendicular to the plane of the sky,
the small clump arrived from the NE, merged with the large clump, and
now the gas from the small clump is located in the SW region, with
respect to the cluster emission center.

In the scenario that radio relics should be recently triggered by a
shock wave for an efficient particles acceleration mechanism, the
projection effects could help us to reconcile the relic position with 
the  apparent position of the temperature enhancement. The A548b 
complex scenario, i.e. the relic's location (on the same
side with respect to the cluster center) and the shock just before the
relic's location, can be explained by the merger geometry sketched in
Fig. \ref{projection}. The two radio relics A and B, located
at projected distances of 500 and 430 kpc from the cluster center, lie
at the boundary of the X-ray brightness distribution, but this is due to
strong projection effects of a merger along the line of sight. 

The two relics represent two enhanced regions belonging to a large-scale 
shock, originated at the cluster center and moving to the peripheral region.
A second symmetrical shock is expected by simulations, but because of mass
ratio and merging geometry, it should be present in a 
peripheral region in SW direction (see Fig. 13) with a very low density.
It cannot be present in our observation because of the chosen pointing center, 
but we do not expect that it could also be seen in a pointed XMM observation.
According to present models of the formation of radio relics, we can imagine 
that a bubble of relativistic particles  is accelerated by the shock originated
by the major merger, the local magnetic field is compressed and amplified,
originating the radio emission. The origin of relativistic particles is
unknown. We note that relativistic particles should not be present in the whole
shocked region, but only in the region where we see the extended radio 
emission. This suggests that these relativistic particles, possibly originating in past AGN activity, were confined in floating discrete bubble regions.

In this scheme, we observe the shock before the relic location due to
the projection effects, because the merger collision is nearly
perpendicular to the plane of the sky.  Due to the poor statistics,
the temperature in the relic regions is poorly constrained, but we
expect these regions also to be shocked regions.  In this scenario,
the relics are at the same distance from the cluster center as the
detected hot region, although the projected distance is smaller.  The
origin of relativistic particles responsible for the relic radio
emission is puzzling. Indeed, it seems that the relativistic particles
are not present in the fully shocked region, but only where radio emission is
detected.

\section{Conclusions}

The main results of our study of the galaxy cluster Abell 
548b from XMM-Newton observations can be summarized 
as follows: 

$X-ray$ $analysis.$
We used XMM-Newton data to obtain X-Ray images and temperature distribution
of the thermal ICM in A548b.
We found that the $\beta$-model is not
a good fit of the data for the whole radial surface brightness 
profile in different sectors towards the N, S, W 
and E. Moreover we observed significant perturbations 
at 1.5$\arcmin$, 5$\arcmin$, and 7$\arcmin$ from the cluster center in all the surface brightness 
profiles we obtained in different sectors and directions. The  
most significant perturbation is present towards relics A and B. 
Perturbations are also present in the residual image after 
subtraction of a 2D $\beta$ model, and here the most relevant 
perturbations are in the region towards the relics. 

We studied the dynamics of A548b from a detailed spectral analysis. 
The central cluster temperature estimated in the inner 3$\arcmin$ is 
3.4$\pm$ 0.1 keV. The spectral study confirms that this cluster is not 
relaxed. We obtained a radial temperature profile, 
which shows a clear temperature increase at 5$\arcmin$ from 
the cluster center. We extracted spectra in the 
regions between the cluster center and the relics A and B, 
and found that the most significant increase in temperature is
present in the N (in direction of the two relics),
but definitely before 
the relic location. We studied the temperature distribution 
in sectors, in boxes, and with the X-ray wavelet spectral 
mapping algorithm. In all cases we obtained similar results: a temperature increase in the annulus 
at 5$\arcmin$ from the cluster center. 
The most significant increase in the temperature (from 
3.5 keV to 6 keV) was observed towards the relic's direction. 
The temperature distribution shows a similar asymmetry to the brightness distribution.on.

$Optical$ $analysis.$
From an optical analysis we observed a large galaxy velocity dispersion of
 $\sigma _V$ $\simeq$ 1300kms$^{-1}$, which is likely to be derived from two groups of galaxies with different 
velocities and velocity dispersions:
``red''-$\sigma_V \simeq~935$kms$^{-1}$ and $z=0.0432$;
``blue''- $\sigma_V\simeq~664$kms$^{-1}$ and $z=0.038$. Assuming that
the ``red'' galaxies are associated with the cluster A548b, the ``blue'' 
galaxies are associated with a merging  clump, with a 1:2-3 mass ratio.
We can explain the apparent large $\sigma_V$ and the
spatial distribution of the ``blue'' galaxies if we consider the
following merging scheme.
Since A548b is in a rich environment with two other clusters A548-2 and A548a, 
it would be the merger of two units, a major component
and a smaller one coming from the NE (from A548a).
From the optical data we know, that A548a, A548b, and A548-2
are located at redshift 0.0394, 0.0423, and 0.0429, respectively.
In this scheme, the collision axis forms a relatively large angle with the plane of the sky. 

$Simulation$ $analysis.$
We used Cluster 6, obtained from the hydro N-body cosmology simulation, code 
RAMSES (\citealt{Teyssier}), to draw the dynamical history of A548b.
The mean temperature obtained from simulations is 3.3 keV, consistent 
with the temperature observed in A548b. We compared simulations and X-ray data for 
density, temperature, and Mach number maps 
at epochs around the maximum core collapse, in particular
just after the maximum core collapse. 
We also studied the influence of projection effects on the observations. 
After this simulation analysis, we selected the output where
density and temperature simulated maps were very similar to
images obtained from our observation.
The better agreeement was found under Y projection, i.e. with the merger 
axis nearly perpendicular to the plane of the sky, and at the epoch
just after the first maximum core collapse.
To compare our results with the radio data, we 
also derived the Mach number map from the previous simulations. 

$X-ray$ $and$ $radio$ $connection.$ A548b is a cluster in a complex
dynamical state. 
From the current data, it is not possible to establish if the central relic C is associated with 
any shock or X-ray feature. Higher angular resolution data are
needed to  analyze this cluster region. 
We observed an increase in the temperature at 5$\arcmin$ -7$\arcmin$ from the cluster center, i.e. before the radio relics A and B. This is interpreted as being caused by a shock 
with Mach number $\sim$1.5. We can explain the relative 
location of the relics with respect to the shocks by strong projection 
effects due to a merger occurring nearly perpendicularly to 
the plane of the sky (see Fig. \ref{projection}). This scenario 
agrees with the suggestion that relics A and B could be part 
of a common structure (\citealt{Feretti2006}). 

\begin{acknowledgements}
We would like to thank Franco Vazza for useful discussions.
R.T. would like to thank Stefan Gottloeber for providing the Cluster 6 initial conditions.
The simulation work was done with the support of the Horizon project.
This research also made use of the NASA/IPAC Extragalactic Database (NED), which is operated by the JPL, CalTech, under contract with NASA.
\end{acknowledgements}


\begin{thebibliography}{}

     \bibitem[Arnaud et al. (2001)]{Arnaud2001} Arnaud, M., Neumann, D.M., Aghanim, N., et al, 2001,
     A\&A, 365, L80

    \bibitem[Arnaud et al. (2002)]{Arnaud2002} Arnaud, M.,Majerowicz, S., Lumb D., et al 2002,
      A\&A, 390, 27

    \bibitem[Belsole et al. (2005)]{Belsole2005} Belsole, E., Sauvageot, J.L., Pratt, G.W., \& Bourdin H. 2005,
     A\&A, 430, 385

    \bibitem[Berrington et al. (2005)]{Berrington2005} Berrington, Robert C., Lugger, P. M., Cohn, H. N., 2002,
     AJ, 123, 2261

    \bibitem[Bohringer et al. (2004)]{Bohringer2004} Bohringer, H., Schuecker, P., Guzzo, et al 2004,
     A\&A, 425, 367

      \bibitem[Briel et al. (2004)]{Briel2004} Briel, U. G., Finoguenov, A.,  Henry, J. P., 2004,
     A\&A, 426, 1

    \bibitem[Cavaliere and Fusco-Femiano (1976)]{Cavaliere1976} Cavaliere, A., Fusco-Femiano, R. 1976,  A\&A, 49, 137

     \bibitem[Chatzikos et al. (2006)]{Chatzikos2006} Chatzikos, M., Sarazin, C. L., Kempner, J. C., 2006,
     ApJ, 643, 751

    \bibitem[Clarke and  Ensslin (2006)]{Clarke2006} Clarke, T. E., Ensslin, T. A.,  2006,
      AJ, 131, 2900

    \bibitem[Ensslin and  Bruggen (2002)]{Ensslin2002} Ensslin, T. A., Bruggen, M., 2002,
     MNRAS, 331,1011


    \bibitem[Davis et al. (1995)]{Davis1995} Davis, D., Bird, C., Mushotzky, R., Odewahn, S.,  1995,
     ApJ, 440, 48

     \bibitem[Dolag (2006)]{Dolag2006} Dolag, K., 2006, AN, 327, 575

    \bibitem[Feretti et al. (2006)]{Feretti2006} Feretti, L., Bacchi, M., Slee, O. B., Giovannini, G., Govoni, F.,      Andernach, H., Tsarevsky, G., 2006,
    MNRAS, 368, 544

    \bibitem[Feretti and Neumann (2006)]{Feretti2006a} Feretti, L., Neumann, D. M.,  2006,
     A\&A, 450L, 21

    \bibitem[Feretti (2003)]{Feretti2003}  Feretti, L.,  2003,
     ASPC, 301, 143

    \bibitem[Feretti \& Giovannini (2007)]{Feretti2007}  Feretti, L., 
Giovannini, G., 2007, Springer Lecture Notes in Physics, Ed. M. Plionis 
et al., in press (arXiv:astro-ph/0703494)

     \bibitem[Ferrari (2005)]{Ferrari2005} Ferrari, C., Benoist, C., Maurogordato, S.,  et al, 2005,
     A\&A, 430, 19

   \bibitem[Ferrari (2006)]{Ferrari2006}  Ferrari, C., Arnaud, M., Ettori, S., Maurogordato, S., Rho, J.,
    A\&A, 446, 417

 \bibitem[Fujita (2006)]{Fujita2006} Fujita, Y., Sarazin, C. L., Sivakoff, G. R., 2006,
 PASJ, 58, 131

    \bibitem[Giovannini \& Feretti (2000)]{Feretti2000} Giovannini, G. , Feretti, L., 2000,
      New Astronomy 5, 335

    \bibitem[Giovannini \& Feretti (2004)]{Feretti2004} Giovannini, G. , Feretti, L., 2004, JKAS 37, 323


      \bibitem[Govoni \& Feretti (2004)]{Govoni2004} Govoni, F., Feretti, L., 2004,
      IJMPD, 13, 1549

    \bibitem[Henriksen et al. (2004)]{Henriksen2004} Henriksen, M., Hudson, D.,
     2004, JKAS, 37, 299

      \bibitem[Henry et al. (2004)]{Henry2004} Henry, J. P., Finoguenov, A., Briel, U. G., 2004,
      ApJ, 615, 181

    \bibitem[Hoeft et al. (2004)]{Hoeft2004}  Hoeft, M., Brggen, M., Yepes, G.,
     2004, MNRAS, 347, 389

    \bibitem[Kempner et al. (2003)]{Kempner2003} Kempner, J., Sarazin, C., Markevitch, M.,  2003,
     ApJ, 593, 291

    \bibitem[Kempner and David (2004)]{Kempner2004} Kempner, J. C., David, L. P., 2004,
     MNRAS, 349, 385

     \bibitem[Landau \& Lifshitz (1959) ]{Landau1959} Landau, L. D., \& Lifshitz, E.M., 1959,
     Fluid Dynamics
     
       \bibitem[Lumb et al. (2002)]{Lumb} Lumb, D.H., Warwick, R.S., Page, M., and De Luca, A., 2002, A\&A
    389, 93-105

     \bibitem[Markevitch et al. (2000)]{Markevitch2000} Markevitch, M., Ponman, T. J.,Nulsen, P. E. J., et al, 2000, ApJ, 541, 542

    \bibitem[Markevitch et al. (2001)]{Markevitch2001} Markevitch, M., Vikhlinin, A.,  2001,
     ApJ, 563, 95
     
     \bibitem[Markevitch \& Vikhlinin (2007)]{Markevitch2007} Markevitch, M., Vikhlinin, A.,  2007,
    Phys. Rep., 443, 1

    \bibitem[Majerowicz et al. (2002)]{Majerowicz2002} Majerowicz, S., Neumann D. \& Reiprich, T.  2002,
      A\&A, 394, 77

    \bibitem[Majerowicz et al. (2004)]{Majerowicz2004} Majerowicz, S., Neumann, D.M., Romer, A.K., et al 2004,
      A\&A, 444, 673

    \bibitem[Mazzotta et al. (2002)]{Mazzotta2002} Mazzotta, P., Fusco-Femiano, R., Vikhlinin, A., Markevich, M, 2002,
    ApJ, 569L, 31

    \bibitem[Mazzotta et al. (2004)]{Mazzotta2004} Mazzotta, P., Brunetti, G., Giacintucci, S., Venturi, T., Bardelli, S., 2004,
      JKAS, 37, 381
      
     \bibitem[Neumann (2005)]{Neumann2005} Neumann, D., 2005,
     A\&A, 439, 465 

    \bibitem[Neumann \& Arnaud (1999)]{Neumann1999} Neumann, D. \& Arnaud M. 1999,
    A\&A, 348, 711

    \bibitem[Nevalainen et al. (2005)]{Nevalainen2005} Nevalainen, J., Markevitch, M., Lumb, D. 2005,
    ApJ, 629,172

    \bibitem[Ricker and Sarazin (2001)]{Sarazin2001} Ricker, P.M.  and Sarazin, G.L. 2001,
    ApJ, 561, 621

    \bibitem[Sauvageot et al. (2005)]{Sauvageot2005} Sauvageot, J.L., Belsole, E.,\& Pratt, G.W. 2005,
      A\&A, 444, 673

     \bibitem[Schindler (2002)]{Schindler2002} Schindler, S., Mergers of Galaxy Clusters in Numerical Simulations, 2002, 229
     
     \bibitem[Solovyeva et al. (2007)]{Solovyeva2007} Solovyeva, L., Anokhin, S., Sauvageot, J. L., Teyssier, R., Neumann, D. 2007,
      A\&A, 476, 63

    \bibitem[Vazza, Brunetti \& Gheller 2008]{Vazza2008} Vazza F., Brunetti G. \& Gheller C., 2008,
     submitted to MNRAS

    \bibitem[Venturi et al. (2007)]{Venturi2007} Venturi, T., Giacintucci, S., Brunetti, G.,
     Cassano, R., Bardelli, S., Dallacasa, D., Setti, G., 2007,
     A\&A, 463, 937

     \bibitem[Vikhlinin et al. (2001)]{Vikhlinin2001} Vikhlinin, A., Markevitch, M., Murray, S. S., 2001,
     ApJ, 551, 160V

     \bibitem[Teyssier (2002)]{Teyssier} Teyssier, R., 2002,
     A\&A 385, 337

    \bibitem[White (2000)]{White2000} White, D.A., 2000,
     MNRAS, 312, 663

    \bibitem[White et al. (1997)]{White1997} White, D.A., Jones, C., Forman, W., 1997,
    MNRAS, 292, 419
    
    \bibitem[Xue et al. (2000)]{Xue2000} Xue, Y.-J., Wu, X.-P., 2000,
     ApJ, 538, 65X

\end{thebibliography}
\end{document}